\documentclass[11pt]{article}
\usepackage{amssymb,cite,amsmath,graphicx,epsf}
\usepackage{epstopdf}
\usepackage[totalwidth=450pt,totalheight=584pt]{geometry}
\newcommand\blank[1]{}

\newcommand{\fract}[2]{{\textstyle\frac{#1}{#2}}}

\newcommand{\eps}{\varepsilon}

\newcommand\RR{{\mathbb R}}

\newcommand{\balpha}{\alpha\kern -6.7pt\alpha}
\newcommand{\bbalpha}{\alpha\kern -4.95pt\alpha}

\newcommand{\CQ}{Q}

\newcommand{\I}{A}

\newcommand\eq{\begin{equation}}
\newcommand\en{\end{equation}}
\newcommand\bea{\begin{eqnarray}}
\newcommand\eea{\end{eqnarray}}
\newcommand\nn{\nonumber}

\newcommand{\One}{{\hbox{{\rm 1{\hbox to 1.5pt{\hss\rm1}}}}}}
\renewcommand{\One}{{\mathbb 1}}
\renewcommand{\One}{{\rm 1\!\!1}}

\newcommand{\Ad}{{\text{AdS}_4/\text{CFT}_3}}
\newcommand{\Bd}{{\text{AdS}_5/\text{CFT}_4}}

\newcommand{\bgammax}{\bar{\gamma}_{\sf x}}
\newcommand{\bgammao}{\bar{\gamma}_{\sf o}}

\newcommand{\RRe}{\text{Re}}
\newcommand{\IIm}{\text{Im}}
\newcommand{\ba}{\begin{eqnarray}}
\newcommand{\ea}{\end{eqnarray}}
\newcommand{\vep}{\varepsilon}

\newcommand{\be}{\begin{equation}}
\newcommand{\ee}{\end{equation}}

\newcommand{\CJ}{P}

\newcommand{\La}{\Lambda}

\newcommand{\z}{s}
\newcommand{\zi}{t}

\renewcommand{\&}{& {\!\!\!\!\!\!}}
\newcommand{\bbT}{\mathbb{T}}
\newcommand{\bfT}{{ \bf T }}
\newcommand{\bbhat}{\hat{\mathbb{T}}}
\newcommand{\bfhat}{\hat{{ \bf T }}}
\newcommand{\barBB}{\underline{\hat{\mathbb{T}}}}

\newcommand{\mF}{\mathcal{F}}
\newcommand{\circl}{{\circlearrowleft}}
\newcommand{\circr}{{\circlearrowright}}

\begin{document}
\newcommand{\figQ}{{\includegraphics[scale=0.5]{ico_m.eps}}}
\newcommand{\figyp}{{\includegraphics[scale=0.5]{ico_f1.eps}}}
\newcommand{\figym}{{\includegraphics[scale=0.5]{ico_f2.eps}}}
\newcommand{\figw}{{\includegraphics[scale=0.5]{ico_b.eps}}}
\newcommand{\figv}{{\includegraphics[scale=0.5]{ico_p.eps}}}
\newcommand{\YYs}{{(\ref{Y1I}-\ref{yv})}\;}
\newcommand{\TTBA}{{(\ref{TBA1a}-\ref{TBA4})}\;}
\newcommand{\DDs}{{(\ref{d0}-\ref{d1})}\;}

\begin{titlepage}
\vskip 2.8cm
\begin{center}
{\Large {\bf Discontinuity relations for the $\text{AdS}_4/\text{CFT}_3$ correspondence} } \\[5pt]
{\Large {\bf } }
\end{center}
\vskip 0.8cm

\centerline{Andrea Cavagli\`a%
\footnote{{\tt Andrea.Cavaglia.1@city.ac.uk}},
Davide Fioravanti%
\footnote{{\tt Fioravanti@bo.infn.it}}
and Roberto Tateo%
\footnote{{\tt Tateo@to.infn.it}}}
\vskip 0.9cm
\centerline{${}^{1,3}$\sl\small Dip.\ di Fisica
and INFN, Universit\`a di Torino,} \centerline{\sl\small Via P.\
Giuria 1, 10125 Torino, Italy}
\centerline{${}^{1}$\sl\small
Centre for Mathematical Science, City University London,}\centerline{\sl\small
Northampton Square, London EC1V 0HB, UK}
\centerline{${}^{2}$\sl\small INFN-Bologna and Dipartimento di Fisica e Astronomia,
Universit\`a di Bologna,} \centerline{\sl\small Via Irnerio 46, 40126 Bologna, Italy}
\vskip 1.25cm
\vskip 0.9cm
\begin{abstract}
\vskip0.15cm
We study in detail the analytic properties of the  Thermodynamic Bethe Ansatz  (TBA) equations for  the anomalous 
dimensions of composite operators in  
the planar limit of the 3D $\mathcal{N}=6$ superconformal Chern-Simons gauge theory
and derive functional relations for the jump discontinuities across the branch cuts in the complex rapidity plane. 
These relations encode the analytic structure of the Y functions and are extremely similar to the ones obtained for
the previously-studied $\Bd$ case. Together with the Y-system and more basic analyticity conditions, they are completely equivalent to the TBA equations.  
We expect these results to be useful to derive alternative nonlinear integral equations for the $\Ad$ spectrum.
\end{abstract}
\end{titlepage}
\setcounter{footnote}{0}
\newcommand{\resection}[1]{\setcounter{equation}{0}\section{#1}}
\newcommand{\appsection}[1]{\addtocounter{section}{1}
\setcounter{equation}{0} \section*{Appendix \Alph{section}~~#1}}
\renewcommand{\theequation}{\thesection.\arabic{equation}}
\def\thefootnote{\fnsymbol{footnote}}
%
%
\resection{Introduction}
In recent years, the research in high energy theoretical physics  has been characterized  by the discovery  
of deep connections between strings, supersymmetric gauge theories and integrable models.
A first link between a  quantum integrable model and multicolor reggeised gluon scattering  was  discovered by 
Lipatov in~\cite{L}, see also~\cite{Faddeev}. More recently, the methods of integrability have turned out to be very efficient for the
 study of some prominent examples of string/gauge theories related by the AdS/CFT correspondence~\cite{Malda,Gubser,Witten}.
For a review of the rapidly developing field of integrability in AdS/CFT the reader can consult \cite{Beisert:review}.

In this paper we are concerned with the Thermodynamic Bethe Ansatz (TBA) approach to the computation of the planar spectrum of the $\Ad$
correspondence \cite{Aharony:2008ug}. This is the spectrum of the anomalous dimensions of local gauge invariant operators in the $\mathcal{N}=6$
superconformal 
Chern-Simons gauge theory, or, equivalently, the energy spectrum of Type IIA string theory on $AdS_4 \times \mathbb{CP}^4$. 
The development of this subject has been parallel and inspired by the study of the spectrum of the 
$\Bd$ correspondence, and we will often refer to the latter. In fact, it is in the context $\Bd$ that integrability was originally discovered.

In the context of $\Ad$, Asymptotic Bethe Ansatz (ABA) equations for anomalous dimensions of composite trace operators were proposed in a series
of seminal works \cite{MinahanAdS4,BS,GromovABA,Ahn}, however a crucial limitation has soon emerged as a consequence of the  
asymptotic character of these equations: the BA equations  do not contain information on  the 
finite size contributions that appear when the site-to-site interaction range in the loop expansion of the dilatation operator becomes greater
than the number of elementary operators in  the trace.

Although, for supersymmetry reasons,  these wrapping effects~\cite{Sieg:2005kd, Plefka, AJK}  do not affect  special families of (protected) operators,
in general  these corrections  become  particularly relevant in the semiclassical limit of string theory corresponding  to  the strong coupling regime on the gauge  theory side. 

This limitation can be surmounted through the use of the Thermodynamic Bethe Ansatz technique \cite{YY}; a  method originally  proposed by   
Al.B. Zamolodchikov~\cite{Zamolodchikov:1989cf}  to
study the ground state energy of perturbed conformal field theories on a cylinder geometry using  the exact knowledge of the scattering data.  
The  method  was later adapted  to the study of excited states~\cite{Bazhanov:1996aq, Dorey:1996re, Dorey:1997rb}.
 The use of the TBA method to overcome the wrapping problem in AdS/CFT was advocated in \cite{AJK} and implemented 
in~\cite{GKVII, Bombardelli:2009ns, Arutyunov:2009ur} for the $\Bd$ case, and in \cite{AdS4,Levkovich} for the $\Ad$ case (see also the review 
\cite{Bajnok}). 
As a result  of this  procedure, the  value of anomalous dimensions as function of the coupling constant is represented in
terms of  the  pseudoenergies  $\varepsilon_a$, solutions of a set of nonlinear coupled integral  equations: the TBA equations. 
Starting from the latter, sets of finite difference  functional relations for  
$Y_a=e^{\varepsilon_a}$, the Y-systems~\cite{Kirillov, Zamolodchikov:1991et, Kuniba:1992ev, Ravanini:1992fi}, 
have been derived for the $\Ad$ and $\Bd$ spectra in ~\cite{Bombardelli:2009ns, GKVII, Arutyunov:2009ur, AdS4,Levkovich}. Apart for a subtle 
small difference crucial  to describe certain  subsectors of the $\Ad$ theory, the  earlier proposal by Gromov, Kazakov and Vieira 
coming from  symmetry arguments~\cite{GKVI} were confirmed.

Y-systems are currently  playing an important r\^ole in Cluster Algebra, gluon scattering amplitudes and other areas of mathematical 
physics \cite{Kuniba:review}. They are related to discrete Hirota's  equations  and they are central in the TBA setup since 
they  exhibit  a very   high degree of universality:  
the whole set of excited states of a given theory is associated  to the same   Y-system 
with different states differing  only in the   number and positions  of the $1+Y_a$ zeros in a certain  fundamental strip. In principle, one can then obtain TBA equations describing the excited states by making natural assumptions on the position of these zeroes and reconstructing the TBA from the Y-system. Excited state TBA equations have been conjectured only for particular subsectors of the AdS/CFT theories and studied in~\cite{GKVII, Gromov:Konishi, Arutyunov:Exploring, Arutyunov:Exceptional, Levkovich, Levkovich:Numerical}. 

In relativistic-invariant  models the Y functions are, in general,  meromorphic  in the rapidity $u$ with zeros and poles both linked to  $1+Y_a$ 
zeros through  the Y-system. 
The situation for the AdS/CFT-related models is further complicated by the presence of square root branch cuts inside and at the 
border of a certain  fundamental strip. According to the known  Y-system to TBA transformation procedures, full information on the Y function 
jump discontinuities across these closest cuts should be independently supplied. 
The main goal of the paper \cite{Extended} was to show that, for the $\Bd$ case, the relevant analytic structure can be  encoded in 
the Y-system together with a set of state-independent functional relations involving points on different 
Riemann sheets.  
Indeed, the results of \cite{Extended} turned out to be very useful both for finding new families of excited states \cite{Balog:disco} and for  
the derivation  of important alternative  non-linear integral equations \cite{Gromov:Solving} for the anomalous dimensions.

In this paper we shall discuss the  $\Ad$ case using a somehow  complementary approach: while in  \cite{Extended} it was shown in detail how 
to trasform the functional relations to the TBA form, here we will start from the $\Ad$ TBA and describe, in reasonable detail,  
how to extract  the full set of discontinuity relations.  A useful additional identity for the fermionic nodes is derived carefully
 in Appendix \ref{additional} showing that it is a consequence of the fundamental discontinuity relations and of the Y-system.

The paper is organised as follows. Section~\ref{sec1} contains
the TBA equations  of~\cite{AdS4, Levkovich}. The Y-system and the new functional relations are presented in Section \ref{sec2}.
In Section \ref{dir}, we provide a concise derivation of these relations from the TBA equations, while in Section \ref{graphical} we show 
how to generate, from the standard Y-system, other non fundamental identities describing the branch cuts far from the real axis.

There are four Appendices. Appendix \ref{kernels} contains a list of the kernels entering the TBA equations, in Appendix \ref{additional} we 
derive a useful additional relation for the fermionic nodes, and in Appendix \ref{list} we list other identities, which can be used to check 
that the Y-system supplemented by the branch cut information is equivalent to the TBA. Finally, in Appendix \ref{appT}, we rewrite the 
fundamental set of relations in terms of the T functions, connecting with the results of \cite{Balog:disco,Gromov:Solving}.

\resection{The TBA equations}\label{sec1}
\begin{figure}
\begin{minipage}[b]{0.5\linewidth}
\centering
\includegraphics[width=6cm]{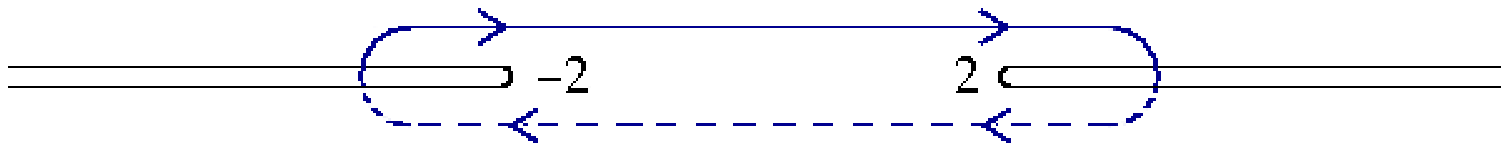}
\caption{The contour $\bgammao$.}
\label{gM}
\end{minipage}
\hspace{0.3cm}
\begin{minipage}[b]{0.5\linewidth}
\centering
\includegraphics[width=6cm]{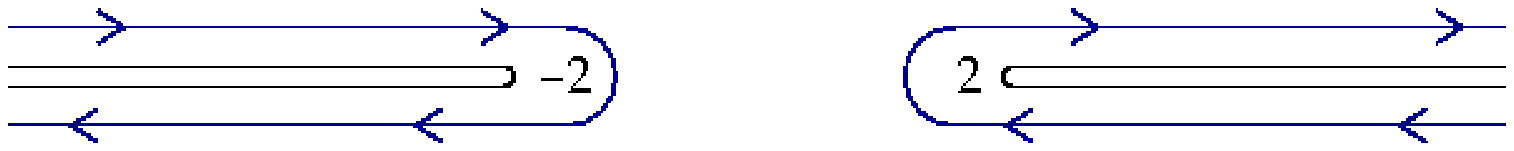}
\caption{The contour $\bgammax$.}
\label{gD}
\end{minipage}
\end{figure}
The TBA equations for the spectrum of $\Ad$ are an infinite set of coupled nonlinear integral equations, depending explicitly on an integer 
parameter $L$ related to the number of elementary fields in the composite operator under consideration, and depending on the 
coupling\footnote{{ In the case of $\Ad$, the coupling $h$ entering the S-matrix elements is a so far undetermined function of the t'Hooft coupling 
$h = h(\lambda)$. }} $h$ through the form of the integral kernels $\phi_{a, b}$. 
Their solutions are a set of pseudoenergies $\eps_a(u)$, of the following species:
$\eps_{\CQ | I}(u)$, $\eps_{\CQ|II}(u)$, $\eps_{w|M}(u)$, $\eps_{v|N}(u)$, $\eps_{y|-}(u)$, $\eps_{y|+}(u)$, where $\CQ, M , N \in \mathbb{N}$.

It is often useful to consider the so-called Y functions, obtained by exponentiating the pseudoenergies:
$Y_a(u)=e^{\vep_a(u)}$. We also set
\be
 L_a(u)= \ln(1+1/Y_a(u)).
\label{LA}
\en
The TBA equations describing the ground state have been derived in \cite{AdS4,Levkovich} and can be written as
\ba
\vep_{\CQ|{\alpha}}(u)&=& \ln{ \lambda_{\CQ|\alpha}} + L\tilde{E}_{\CQ}(u)
-\sum_{\beta}\sum_{\CQ'=1}^{\infty}L_{\CQ'|\beta}*\phi_{(\CQ'|\beta),(\CQ|\alpha)}(u)
+\sum_{M=1}^{\infty}L_{v|M}*\phi_{(v|M),\CQ}(u) \nn \\
&+&
\int_{-2}^{2} dz\left[L_{y|-}(z)\,\phi_{(y|-),\CQ}(z,u)
- L_{y|+}(z)\,\phi_{(y|+),\CQ} (z,u)\right]
~,\label{TBA1a} \\
\vep_{y| \mp }(u)&=& \ln{ \lambda_{y| \mp }} -\sum_{\CQ=1}^{\infty}(L_{\CQ|I }+L_{\CQ|II})*\phi_{\CQ,(y| \mp )}(u)
+ \sum_{M=1}^{\infty} (L_{v|N}-L_{w|M}  )*\phi_{M}(u)~, \label{TBA2}\\
\vep_{v|K}(u)&=& \ln{ \lambda_{v|K} } -
\sum_{\CQ=1}^{\infty}(L_{Q|I}+L_{Q|II})*\phi_{\CQ,(v|K)}(u)
+\sum_{M=1}^{\infty}L_{v|M}*\phi_{M,K}(u) \nn \\
&+& \int_{-2}^{2} dz\,(L_{y|-}(z)-L_{y|+}(z))\, \phi_K(z-u)
~,~~~\label{TBA3} \\
\vep_{w|K}(u)&=& \ln{ \lambda_{v|K} } + \sum_{M=1}^{\infty}L_{w|M}*\phi_{M,K}(u)+ \int_{-2}^{2} dz\,(L_{y|-}(z)-L_{y|+}(z))\, \phi_{K}(z-u)
~,
\label{TBA4}
\ea
for $\alpha=I,II$, $K = 1, 2, \dots$, and the fugacities $\lambda_a$ 
will be specified below. The integral kernels $\phi_{a, b}(z, u)$ appearing in the TBA equations are defined in Appendix \ref{kernels}, 
together with the function $\tilde{E}_{\CQ}(u)$, which represents the infinite volume energy of a $\CQ$-particle bound state in the mirror theory.

The ground state energy can be computed as
\eq
 \tilde{F}(L)=-{1 \over L} \sum_{\CQ=1}^{\infty} \int_{\RR} {du \over 2 \pi}  \,  {d\tilde{p}^{\CQ} \over du}  \left( L_{\CQ|I}(u) + L_{\CQ|II}(u)  \right),
\label{FL}
\en
where $\tilde{p}_{\CQ}$ is the mirror momentum, also defined in Appendix \ref{kernels}.
This quantity is exactly zero, as dictated by supersymmetry, as soon as the fugacities reach the values
\eq
\lambda_{\CQ|\alpha}=(-1)^\CQ ~,~~\lambda_{v|K}=\lambda_{w|K}=1~,
~~\lambda_{y|\pm}=-1~,~~~(\alpha =I,II,~K=1,2,\dots)~.
\label{fugacities}
\en
This singular limit of the TBA equations can be regularised by taking
\eq
\lambda_{2Q-1|I}= -e^{i \varphi}~,~~\lambda_{2Q-1|II}= - e^{-i \varphi}~,~~\lambda_{2Q|I}= \lambda_{2Q|II}= 1~,~~\lambda_{v|K}=\lambda_{w|K}=1~,~~\lambda_{y|\pm}=-1,
\label{fugacities1}
\en
such that the TBA equations are regular for $ \varphi \ne 0$ and the ground
state energy tends to zero as $ \varphi\rightarrow 0$. The nontrivial anomalous dimensions corresponding to excited states can be obtained 
considering the TBA equations (\ref{TBA1a}-\ref{TBA4}) and (\ref{FL}) with different integration
contours \cite{Bazhanov:1996aq,Dorey:1996re,Gromov:Konishi,Arutyunov:Exploring}.

There is a crucial difference between this system and the TBA equations describing two dimensional 
relativistic quantum field theories in finite volume: the S-matrix elements listed in 
Appendix \ref{kernels} contain, in addition to poles and zeroes, 
also square root branch points in the rapidity plane. As a consequence, the TBA solutions are multivalued functions with infinitely many 
branch points, whose locations are summarised in Table \ref{table1} for the different Y functions. 
The branch cuts are clearly visible, in the $\Bd$ case, in the numerical study presented in \cite{Analytic}.

\begin{figure}[h]
\centering
\includegraphics[width=9cm]{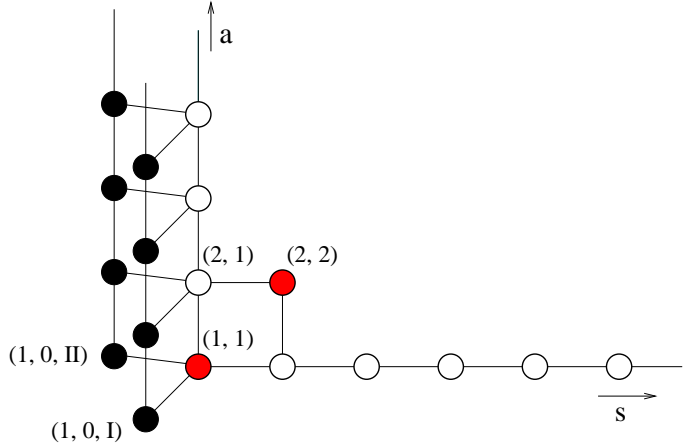}
\caption{The lattice associated to the $\Ad$ Y-system, with
$(\CQ , 0,\alpha) = (\CQ | \alpha)$, $(1,1) = (y|-)$, $(2,2) = (y|+)$, $(n, 1) = (v|n-1)$, $(1,n) = (w|n-1)$. }
\label{figY}
\end{figure}

Let us introduce an important convention which becomes relevant when the solutions to the TBA are continued into the complex rapidity plane.
We work on sections of the Riemann surface obtained 
by tracing every branch cut as a horizontal, semi-infinite segment external to the strip $|\RRe(u)|< 2$. More explicitly, we draw branch cuts of
the form: 
$(-\infty, -2) \cup (+2, +\infty) + i n/h$, where the possible values of $n \in \mathbb{Z}$ are listed in Table \ref{table1}.
Moreover, we denote as the ``first'' Riemann sheet the one containing the physical values of the Y functions on the real axis. 
Whenever we need to reach values of the Y functions on another sheet, we will indicate it explicitly.
\begin{table}[tb]
\begin{center}
\begin{tabular}{|c|cc|}
\hline
Function & Singularity position &\\
\hline
\hline
$Y_y(u)$  & $u=\pm 2 + i \;n_0 /h$, & $n_0 = 0, \pm 2, \pm 4, \dots$\\
\hline
$ Y_{w|M}(u)$ & & \\
\cline{1-1}
$Y_{v|M}(u)$ & $u = \pm 2 + i \; n_{ \small M } /h$, & $ n_{ \small M }= \pm M, \pm (M+2), \pm (M+4),\dots$ \\
\cline{1-1}
$ Y_{M|I}(u),$ $ Y_{M|II}(u)$ & & \\
\hline
\end{tabular}
\caption{\small Square-root branch points for the Y functions.}
\label{table1}
\end{center}
\end{table}
\resection{ The extended Y-system }\label{sec2}
When they are evaluated on the first Riemann sheet, the solutions to the $\Ad$ TBA satisfy the following set of functional relations (the Y-system)
\eq
Y_{1|I}(u+\fract{i}{h}) Y_{1|II}(u-\fract{i}{h})= { \left(1+Y_{2|I}(u)\right) \over
 \left(1+\frac{1}{Y_{y|-}(u)}\right)}, 
\label{Y1I}
\en
\eq
Y_{1|II}(u+\fract{i}{h}) Y_{1|I}(u-\fract{i}{h})= { \left(1+Y_{2|II}(u)\right) \over
 \left(1+\frac{1}{Y_{y|-}(u)}\right)},
\label{Y1II}
\en
\eq
Y_{\CQ|I}(u+\fract{i}{h}) Y_{\CQ|II}(u-\fract{i}{h})= {\left(1+Y_{\CQ-1|II}(u)\right) \left(1+Y_{\CQ+1|I}(u)\right) \over
 \left(1+\frac{1}{Y_{(v|\CQ-1)}(u)}\right)},  \text{  }\CQ = 2, 3, \dots
\label{eq:YQI}
\en
\eq
Y_{\CQ|II}(u+\fract{i}{h}) Y_{\CQ|I}(u-\fract{i}{h})= {\left(1+Y_{\CQ-1|I}(u)\right) \left(1+Y_{\CQ+1|II}(u)\right) \over
 \left(1+\frac{1}{Y_{(v|\CQ-1)}(u)}\right)},  \text{  }\CQ = 2, 3, \dots
\label{eq:YQII}
\en
\eq
Y_{y|-}(u+\fract{i}{h})  Y_{y|-}(u-\fract{i}{h}) = \frac{\left(1+Y_{v|1}(u)\right)}{\left(1+Y_{w|1}(u)\right)}\prod_{\alpha = I, II} 
\left( 1+\frac{1}{Y_{1|\alpha}(u)} \right)^{-1},\label{Yysystem}
\en
\eq
Y_{w|M}(u + \fract{i}{h}) Y_{w|M}(u-\fract{i}{h})=\prod_{N=1}^{\infty}\left(1+Y_{w|N}(u)\right)^{A_{MN}}
\left[\frac{\left(1+\frac{1}{Y_{y|-}(u)}\right)}{\left(1+\frac{1}{Y_{y|+}(u)}\right)}\right]^{\delta_{M, 1}},\label{yw}
\en
\eq
Y_{v|M}(u+\fract{i}{h}) Y_{v|M}(u-\fract{i}{h})=\frac{\prod_{N=1}^{\infty}\left(1+Y_{v|N}(u)\right)^{A_{MN}}}{\left(1+\frac{1}{Y_{M+1}(u)}\right)}
\left[\frac{\left(1+Y_{y|-}(u)\right)}{\left(1+Y_{y|+}(u)\right)}\right]^{\delta_{M, 1}},
\label{yv}
\en
where $A_{MN} = \delta_{M, N+1 } - \delta_{M, N-1}$. The $\Ad$ Y-system has been rigorously derived from the TBA equations in \cite{AdS4,Levkovich},
where the appropriate choice of Riemann section was discussed. In the special symmetric case $Y_{\CQ|I} = Y_{\CQ|II}$, these relations coincide 
with those originally conjectured in \cite{GKVI}. They can be associated to the diagram in Figure \ref{figY}.
Notice that, contrary to the case in relativistic two-dimensional QFTs, where the Y's are meromorphic functions of the rapidity, the Y-system 
contains much less information than the TBA: in fact, it hides away completely the presence of the branch points.
However, analogously to what done in \cite{Extended} for the $\Bd$ Y-system, 
we can encode the missing information in an additional set of local functional relations describing the branching 
properties of the Y functions. 
\begin{figure}[h]
\centering
\includegraphics[width=8cm]{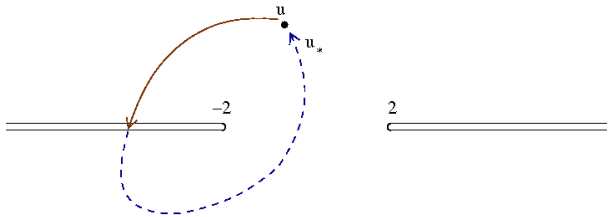}
\caption{ The second sheet image $u_*$ of $u$. }
\label{cutfinal}
\end{figure}
In order to present the result, let us introduce some notation, following \cite{Extended}.  We denote with $u_*$ the image of the point $u$ 
obtained by analytic continuation along the path represented in Figure \ref{cutfinal}, encircling the point $-2$\footnote{ It would be 
completely equivalent to encircle the branch point $+2$, because the topology of the Riemann surface on which the Y functions are defined 
is symmetric under reflection across the imaginary axis. }. Then, let $f(u)$ be a function with a square root branch point at 
$u =  -2 + \frac{i}{h} N $. We describe its local branching properties with a ``discontinuity'' function, denoted by $\left[ f(u) \right]_N $. 
It is defined as the difference between $f( u + \frac{i}{h}N )$ and its image after encircling the branch point $-2 + \frac{i}{h}N $, namely: 
\bea\label{dis}
\left[ f(u) \right]_N = f( u + \frac{i}{h}N ) - f( u_* + \frac{i}{h}N).
\eea
We call it a discontinuity function because, if we restrict to $u \in ( -\infty, -2) \cup (+2, +\infty) + i 0^+$, it returns the value of the jump 
across the branch cut with imaginary part $+ N/h$. However, (\ref{dis}) is more general and defines a new complex function living on a 
multi-sheeted surface. 
Notice that, in general, $\left[ f(u) \right]_N $ itself has infinitely many branch points corresponding to the branch points of $f(u)$.

Finally, let us introduce the following important quantities:
\eq
\Delta^{\alpha}(u)= \left[\ln Y_{1|\alpha}(u)\right]_{+1},  \hspace{0.5cm}\alpha = I, II .
\label{delta0}
\en
We are now ready to write down the extra analytic information that completes the Y-system.
First of all, we assume the knowledge of the position of the branch points inside the \emph{ physical strip } $|\IIm(u)|<1/h$, 
namely the fact that $Y_{y|\pm}(u)$ have two branch points at $u = \pm 2$,
while $Y_{1|w}(u)$, $Y_{1|v}(u)$ and $Y_{1|\alpha}(u)$ ($\alpha = I, II$) have four branch points at $u = \pm 2 + i/h$, $u = \pm 2 - i/h$.

The first fundamental property is that $Y_{y|\pm}$ are two branches of the same function:
\bea
\label{branch}\label{d0}
Y_{y|-}(u_*) = Y_{y|+}( u ).
\eea
The remaining functional relations are:
\eq
\left[\ln{Y_{w|1}} \right]_{1}= \ln\biggl( { 1+1/Y_{y|-} \over  1+1/Y_{y|+}}\biggl),~~~~
\left[\ln{Y_{v|1}} \right]_{1}=
\ln\biggl( { 1+Y_{y|-} \over  1+Y_{y|+} }\biggl),
\label{d3}
\en
\eq
\biggl[\ln\biggl(\frac{Y_{y|-}}{Y_{y|+}}\biggl)\biggr]_{ \pm 2N}=-\sum_{\CQ=1}^N \;
\sum_{\alpha = I, II}\left[\ln\left(1+{1 \over Y_{\CQ|\alpha}}\right)\right]_{\pm (2N-\CQ)},
\label{d2}
\en
\eq
\left[
\Delta^{\alpha} \right]_{\pm 2N}=\mp \biggl[\ln\biggl(1+ \frac{1}{Y_{y|\mp}}\biggr)\biggr]_{\pm 2N}  \mp  
\sum_{M=1}^{N} \biggl[\ln\biggl(1+{1 \over Y_{v|M}}\biggr)\biggr]_{\pm (2N-M)}
    \mp  \ln{\biggl({Y_{y|-} \over Y_{y|+}}\biggl)},\label{d1}
\en
with $\alpha = I, II$, $ N=1,2,\dots$. These relations are extremely similar to the ones appearing in the context of the $\Bd$ correspondence 
and discovered in \cite{Extended}. 
Together with \YYs, they constitute a fundamental set of local and state-independent equations, which is completely equivalent to the TBA.

In the next Section, we show how (\ref{d3}-\ref{d1}) can be extracted from the TBA equations. 
The reconstruction of the TBA from the Y-system equations \YYs {\it extended} by \DDs, conversely, 
is essentially the same as the one contained in \cite{Extended} for the $\Bd$ case and we do not report it here.

As a final comment, notice that there is no dependence on $\alpha$ in the rhs of (\ref{d1}). Therefore, although we expect that 
$ \vep_{1|I}(u) \neq \vep_{1|II}(u)$ for particular 
excited state solutions, the related discontinuity functions are equal, as confirmed by the expression for the ground state (\ref{eq:nonlo}) below. 
In fact, because the two wings $\alpha = I, II$ enter symmetrically in (\ref{eq:nonlo}), any process of analytic continuation will preserve this property,  and we conclude that $\Delta^I(u) = \Delta^{II}(u)$ for any state.
\resection{ A sketch of the derivation }\label{dir}
In this Section we provide a concise derivation of (\ref{d3}-\ref{d1}) from the TBA. 
The idea is to compute the discontinuity functions relative to branch points located in or on the border of the 
physical strip $ | \IIm(u) | < 1/h$, which contains the essential analytic information necessary for the reconstruction of the TBA equations.
 We confine our attention to the following functions: 
$\left[ \ln Y_{w|1}\right]_{ + 1 }$, $\left[ \ln Y_{v|1}\right]_{ + 1 }$, $\left[ \ln Y_{y|-}\right]_{ 0 }$, $\left[ \ln Y_{1|\alpha}\right]_{ + 1 }$ 
($\alpha = I, II$). In fact, functions of the form $\left[ \ln Y_a \right]_{-1}$ can easily be recovered from the ones listed above using the
standard Y-system \YYs.
\subsubsection{ The discontinuity relations for the $v$ and $w$ functions}
\begin{figure}[h]
\centering
\includegraphics[width=11cm]{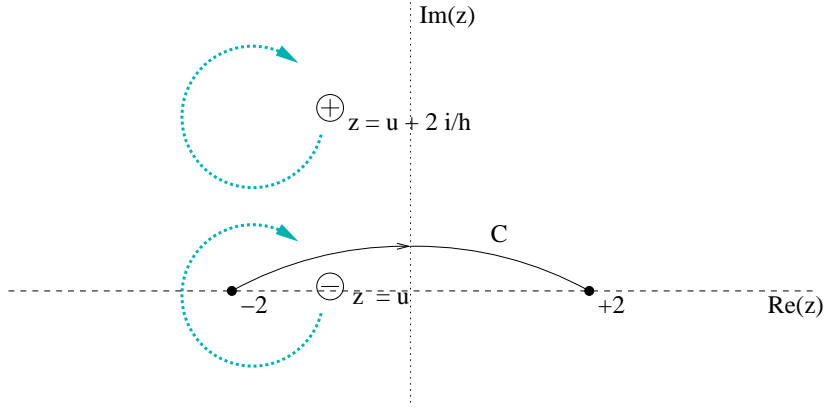}
\caption{An illustration of equations (\ref{expr1}) and (\ref{expr2}). 
The ``plus'' and ``minus'' circles represent the poles of the kernel $\phi_1(z - u)$ 
(with residues $\pm 1$, respectively) after the shift $u \rightarrow u+i/h$, 
while the dashed arrows represent the motion of the poles as a result of $u + i/h \rightarrow u_* + i/h $.}
\label{fig:limconv}
\end{figure}

The easiest quantity to compute is: 
\bea
\label{eq:quantity11}
\left[ \ln Y_{w|1}( u ) \right]_{+1} =\vep_{w|1}( u + \frac{i}{h} ) -  \vep_{w|1}( u_* + \frac{i}{h}  ).
\eea 
We start from the TBA equation
\begin{equation}
\label{contiTBA}
\vep_{w|1}(u)=\sum_{M=1}^{\infty}L_{w|M}*\phi_{M,1}(u)+ \int_{-2}^{2} dz\,(L_{y|-}(z)-L_{y|+}(z))\, \phi_{1}(z-u).
\end{equation}
When analytically continuing equations of TBA type one has to notice that a change in the external variable 
induces a motion of the poles of the integral kernels. Whenever a pole crosses the integration contour on the real axis, 
the form of the equation must be modified, either by deforming the contour or by adding a residue term.
Notice that all the kernels in (\ref{contiTBA}) are meromorphic and therefore no branching can occur as
 long as the poles stay bounded away from the real axis. Considering the kernels in (\ref{contiTBA}), this proves that the 
solution is analytic for $|\IIm(u)|<1/h$. \\
To study the analyticity on the border of this strip we can concentrate on two terms in the rhs of (\ref{contiTBA}): 
the convolution
\bea
\label{eq:triv} 
I_1(u) =  L_{w|2}*\phi_{2,1}(u), 
\eea 
which is potentially dangerous because $\phi_{2, 1}(u) = \phi_{1}(u) + \phi_{3}(u)$ and $\phi_1(z-u)$ has two poles 
at $z = u \pm \frac{i}{h} $, and the integral
\bea
\label{eq:limconv} 
I_2(u) = \int_{-2}^{2} dz\,(L_{y|-}(z)-L_{y|+}(z))\, \phi_{1}(z-u).
\eea
In the case of (\ref{eq:triv}) the integration contour lies on the real axis. With a slight deformation 
it is possible to avoid any contact with the poles, so that this term is analytic on the whole line $\IIm(u) = 1/h$. 
In the case of (\ref{eq:limconv}), deforming the contour we have
\bea\label{expr1}
I_2(u + i/h ) =  \oint_{C} dz\,(L_{y|-}(z)-L_{y|+}(z))\, \phi_{1}(z-u - i/h),
\eea
where the contour $C$ is represented in Figure \ref{fig:limconv}. However, contrary to the previous case, it is 
now impossible to avoid trapping the contour when one of the points $-2$ or $+2$ is encircled. Therefore, taking the residue with the appropriate sign, we find
\bea
\label{expr2}
I_2(u_* + i/h ) &=& I_2( u + i/h) - \left( L_{y|-}(z)-L_{y|+}(z) \right).
\eea
Subtracting (\ref{expr2}) from (\ref{expr1}), we finally find the first relation in (\ref{d3}).
Notice that, for simplicity, we have performed the calculation using a clockwise-oriented path for 
the continuation $u+i/h\rightarrow u_*+i/h$ as shown in Figure \ref{fig:limconv}. 
The reader can check that following an anticlockwise path would lead to the same result 
(therefore showing that the branching is of square-root type) using the property
$L_{y|+}( u_*) = L_{y|-}(u)$, $L_{y|-}( u_*) = L_{y|+}(u)$. 
The fact that $\eps_{y|\pm}(u)$ are branches of the same function is a consequence of the TBA equation (\ref{TBA2}) and of the identity
 (\ref{important}).

The second relation in (\ref{d3}) can be derived exactly in the same way after rewriting the relevant TBA equation 
as\footnote{ We have used equation (\ref{discTBA}) and the simple kernel identity:
\begin{align}
\label{eq:QQ'convol1}
\int_{-2}^{+2} &ds \;\left( K( v - i \CQ/h  , s ) - K( v + i \CQ/h , s ) \right) \phi_M(s-u) + \phi_{\CQ - 1 }(z-u) = \phi_{\CQ, (v|1)}(z, u).
\end{align}
 }
\bea
\vep_{v|1}(u)&=&-
\sum_{\CQ= 2}^{\infty}(L_{Q|I}+L_{Q|II})*\phi_{\CQ - 1 }(u)
+\sum_{M=1}^{\infty}L_{v|M}*\phi_{M,1}(u) \nn\\&+& \int_{-2}^{2} dz\,(\La_{y|-}(z)-\La_{y|+}(z))\, \phi_1(z-u).
\eea
\subsubsection{ The discontinuity relations for the fermionic nodes }
Let us now consider the fermionic excitations $\vep_{ y }$.  
The value of 
\eq
\left[ \ln Y_{y|-}( u ) \right]_{0}  = \vep_{y|-}( u ) - \vep_{y|-}( u_* ) 
\en
can be read from the TBA equations:
\eq
\vep_{y|-}(u) - \vep_{y|+}(u) = -\sum_{\CQ = 1}^{\infty} \int_{-\infty}^{\infty} dv \;( L_{\CQ|I}(v) + L_{\CQ|II}(v) )
\left( K( v - i \CQ/h  , u ) - K( v + i \CQ/h , u ) \right),
\label{discTBA}
\en
where we used the identity (\ref{K1}).

Notice that, contrary to the case of $\left[ \ln Y_{w|1}( u ) \right]_{1}$, $\left[ \ln Y_{v|1}( u ) \right]_{1}$, (\ref{discTBA}) is a non local
expression, and therefore we expect its form to depend 
on the particular excited state under consideration. However, in analogy to what seen in \cite{Extended} in the $\Bd$ context, we can trade
 it for an infinite number of local functional relations describing the discontinuity functions 
$\left[ \ln{ Y_{y|-}(u) / Y_{y|+}(u ) } \right]_{\pm 2N }$, $N = 1, 2, \dots$.
To compute these quantities, it is sufficient to note that, under the analytic continuation 
$u \rightarrow u \pm i 2 N/h$, (\ref{discTBA}) is modified by a number of residue terms. For example if
 $u \rightarrow u + i 2 N/h$, $N \in \mathbb{N}$, with $u < \IIm(u) < 1/h$, $| \RRe(u) | < 2 $, we have
\bea
\label{eq:transport}
&&\ln \frac{ Y_{y|-}(u + i 2N/h) }{Y_{y|+}(u + i 2N/h) } \\
&=& -\sum_{\CQ = 1}^{\infty} \int_{-\infty}^{\infty} dv\;( L_{\CQ|I}(v) + L_{\CQ|II}(v) )
\left( K( v - i \CQ/h  , u + i 2 N/h) - K( v + i \CQ/h , u + 2 i N /h ) \right) \nn\\
&&-\sum_{\CQ = 1}^{ 2 N } (  L_{\CQ|I}( u + i ( 2N - \CQ )/h) + L_{\CQ|II}( u + i ( 2 N - \CQ )/h ) ).
\eea
The analytic continuation $ u + i 2N/h \rightarrow u_* + i 2 N/h  $ has no effect on the convolution in (\ref{eq:transport}) 
and is nontrivial only for half of the residue terms in the last line, because $L_{\CQ|\alpha}(u)$ is analytic for $| \IIm(u)|< \CQ/h$. 
This gives precisely (\ref{d2}).
\subsubsection{ Discontinuity relations for the functions $\Delta^{I} = \Delta^{II}$ }
Finally, let us consider the quantities 
$$\Delta^{\alpha}(u)= \left[\ln Y_{1|\alpha}(u)\right]_{+1} = \vep_{ 1|\alpha}( u + i/h ) - \vep_{ 1|\alpha}( u_* + i/h  ).$$ 
The analytic continuation of the TBA equation (\ref{TBA1a}) leads to the following expression, valid  for $0 < \text{Im}(u) < \frac{2}{h}$: 
\bea
\label{eq:nonlo}
&&\Delta^{I}( u ) = \Delta^{II}(u) \\
&=& L \ln{ x^2(u)} - L_{y|-}(u) + \oint_{\bgammao} ds \; L_y(s) K( s, u ) + \sum_{N=1}^{\infty} \int_{-\infty}^{\infty} 
ds\; L_{v|N}(s) K^{\left\{ N \right\} }(s , u ) \; ds + \Delta^{\Sigma}_4(u)\nn
\eea
where
\eq
K^{\left\{ N \right\} }(s , u ) \equiv K(s + i N/h , u ) + K(s - i N/h , u )
\en
and 
\bea\label{eq:sig4}
\Delta^{\Sigma}_4(u) = \sum_{\alpha = I, II} \sum_{\CQ=1}^{\infty} \int_{-\infty}^{\infty} ds \; L_{\CQ|\alpha}(s) K_{\CQ}^{\Sigma}(s, u) .
\eea
The kernel $K_{\CQ}^{\Sigma}$ sums up the contribution of the dressing factor, appeared already in the $\Bd$ context and is defined by
\bea\label{eq:sig2}
K_{\CQ}^{\Sigma}(s, u) = K_{\CQ, 1}^{\Sigma}(s, u+i/h ) - K_{\CQ, 1}^{\Sigma}(s, u_* + i/h ).
\eea
Using (\ref{dress}), it is easy to show that
\bea\label{sig}
K_{\CQ}^{\Sigma}(s, u) = \oint_{\bgammax} dt \; \phi_{\CQ, y}( s , t ) \oint_{\bgammax} dz \;  K_{\Gamma}^{\left[2\right]} ( t - z ) K( z , u) 
- \oint_{\bgammax} dt \; \phi_{\CQ, y}( s , t ) \;K_{\Gamma}^{\left[2\right]} ( t - u ) .
\eea
Again, equation (\ref{eq:nonlo}) is non-local, and in order to express its analytic content in a state-independent way we consider the following quantities:
\eq
\left[ \Delta^{I} \right]_{\pm 2N}  = \left[ \Delta^{II} \right]_{\pm 2N}, \hspace{0.5cm}N = 1, 2, \dots 
\en
By analytic continuation of (\ref{eq:nonlo}) using the techniques illustrated in the previous paragraphs, we find
\bea\label{eq:finalDSig}
\left[
\Delta^{\alpha} \right]_{\pm 2N}=\mp \left[ L_{y|\mp} \right]_{\pm 2N}  \mp  \sum_{M=1}^{N} \left[L_{v|M} \right]_{\pm (2N-M)} + \left[ \Delta^{\Sigma}_4 \right]_{\pm 2N}, \hspace{0.5cm}\alpha = I, II .
\eea
The last term can be computed explicitly from equations (\ref{eq:sig4} - \ref{sig}). 
Notice that the first convolution on the rhs of (\ref{sig}) has a trivial monodromy for $u$ far from the real axis, 
and therefore we can discard it. On the contrary, applying the sequence of analytic continuations 
$u \rightarrow ( u \pm i 2N/h ) \rightarrow u_* \pm i 2N/h $ for $| \IIm(u) |< 1/h$, $| \RRe(u) |< 2 $, the second convolution in (\ref{sig}) 
transforms as follows:
\bea
&-& \oint_{\bgammax} dt \; \phi_{\CQ, y}( s , t ) \;K_{\Gamma}^{\left[2\right]} ( t - u ) \rightarrow - \oint_{\bgammax} dt \; \phi_{\CQ, y}( s , t ) \;K_{\Gamma}^{\left[2\right]} ( t - ( u \pm i 2 N/h )) \\
&\rightarrow & - \oint_{\bgammax} dt \; \phi_{\CQ, y}( s , t ) \;K_{\Gamma}^{\left[2\right]} ( t - ( u \pm i 2 N/h ) ) \mp \left( \phi_{\CQ, (y|-) }( s , u ) - \phi_{\CQ, (y|-) }( s , u ) \right).\nn
\eea 
Therefore we find
\bea
\label{eq:finalD4}
\left[
\Delta^{\Sigma}_4 (u) \right]_{\pm 2N} &=& \sum_{\alpha = I, II} \sum_{\CQ=1}^{\infty} \int_{-\infty}^{\infty} ds \;
 L_{\CQ|\alpha}(s) \left( K^{\Sigma}_{\CQ}( s, u \pm i 2 N/h ) - K^{\Sigma}_{\CQ}( s, u_* \pm i 2 N/h ) \right )\nn \\
&=& \pm \sum_{\alpha = I, II} \sum_{\CQ=1}^{\infty} \int_{-\infty}^{\infty} ds \; L_{\CQ|\alpha}(s) 
\left(  \phi_{\CQ, (y|-) }( s , u ) - \phi_{\CQ, (y|-) }( s , u ) \right)\nn\\
&=& \mp \ln{( Y_{y|-}(u) / Y_{y|+}(u) )}.
\eea
Combining the latter with (\ref{eq:finalDSig}), we recover (\ref{d1}).
\resection{ More discontinuity relations }\label{graphical}
In this Section we show how to deduce further constraints relating branch points lying inside and outside of the
physical strip, using only the standard Y-system \YYs. As seen 
in \cite{Extended} in the $\Bd$ case, these additional discontinuity relations are useful for the purpose of deriving the 
TBA from the extended Y-system. 
 
\begin{figure}[h]
\begin{minipage}[t]{0.5\linewidth}
\centering
\includegraphics[width=5cm]{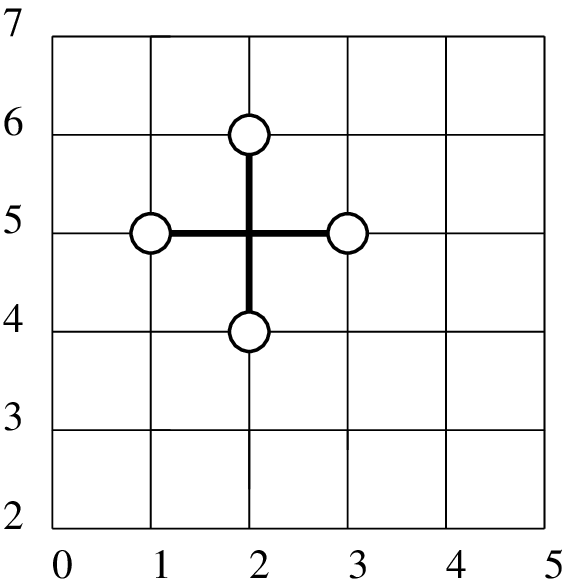}
\caption{ An illustration of relation (\ref{eq:Y2}) for $M=2$ and $K=5$.}
\label{grid}
\end{minipage}
\hspace{0.3cm}
\begin{minipage}[t]{0.5\linewidth}
\centering
\includegraphics[width=5cm]{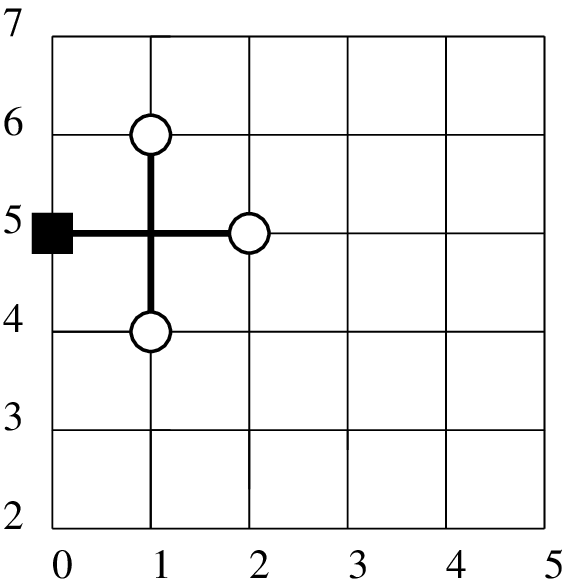}
\caption{ An illustration of relation (\ref{eq:Y2}) for $M=1$ and $K=5$.}
\label{grid2}
\end{minipage}
\end{figure}

\begin{figure}[h]
\begin{minipage}[t]{0.5\linewidth}
\centering
\includegraphics[width=6cm]{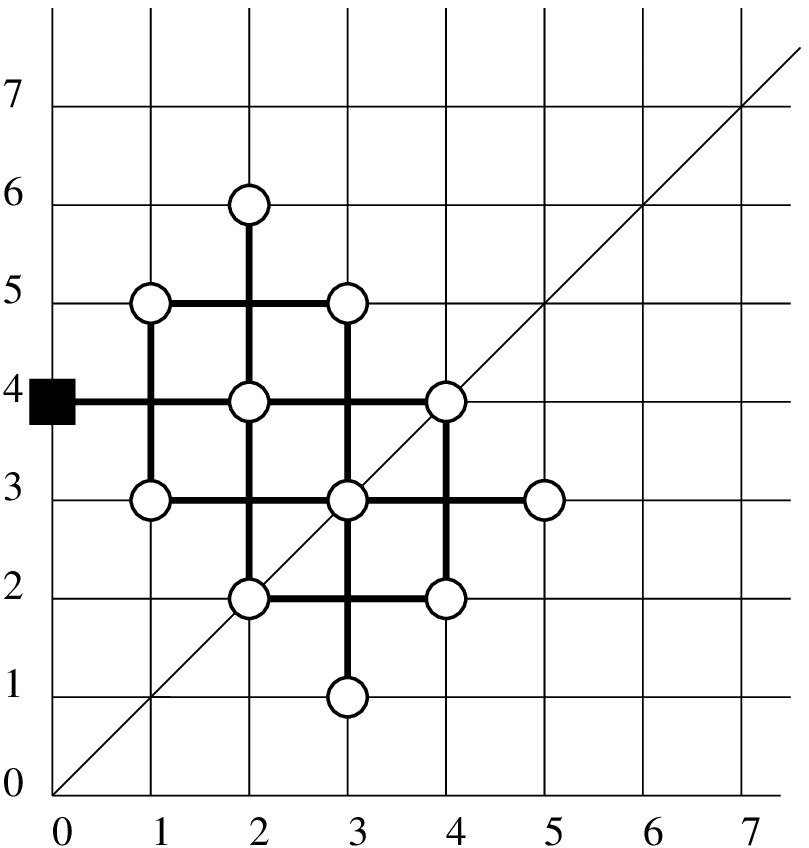}
\caption{ A representation of equation (\ref{eq:def2}) with $N = 2$.}
\label{llow}
\end{minipage}
\hspace{0.3cm}
\begin{minipage}[t]{0.5\linewidth}
\centering
\includegraphics[width=6cm]{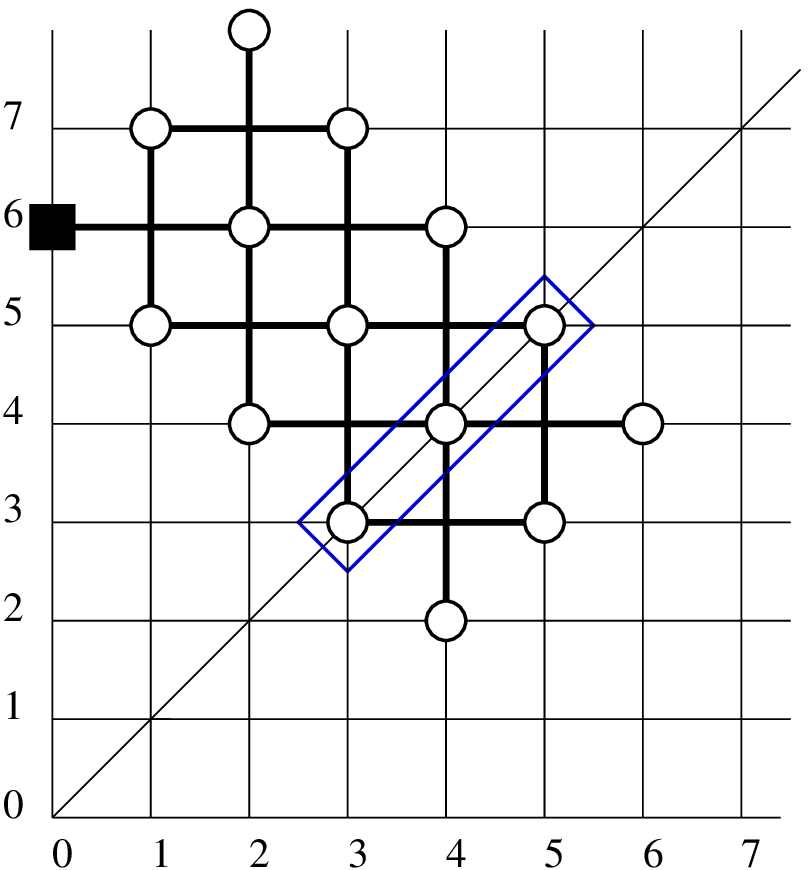}
\caption{ A representation of equation (\ref{eq:def2}) with $N = 3$. The nodes on the diagonal correspond to the rhs of equation (\ref{eq:rulerulew}).}
\label{hhigh}
\end{minipage}
\end{figure}

We illustrate the strategy in the case of the $Y_{M|w}$ functions. From the Y-system equation
\bea\label{eq:Y2}
&&\ln Y_{M|w}( u + i (K+1) /h )  + \ln Y_{M|w}( u + i (K-1)/h ) \\
&=& ( 1 - \delta_{M, 1})\La_{M-1|w}( u + i K/h ) + \La_{ M+1| w }( u + i K/h ) + \delta_{M, 1}
\left( L_{y|-}( u + i K/h ) - L_{y|+}( u + i K/h ) \right)\nn
\eea
it follows that
\bea
\label{eq:YD2}
&&\left[ \ln Y_{M|w}( u ) \right]_{K+1}  + \left[ \ln Y_{M|w}( u ) \right]_{K-1} \\
&=& ( 1 - \delta_{M, 1})\left[ \La_{M-1|w}( u )\right]_K + \left[\La_{ M+1| w }( u  )\right]_K + \delta_{M, 1} 
\left[ \left( L_{y|-}( u  ) - L_{y|+}( u ) \right) \right]_{K},\nn
\eea
for $K \in \mathbb{Z}$. To derive more complicated identities, it is convenient to introduce a pictorial notation. We represent a shifted 
Y-system equation such as (\ref{eq:Y2}) for $M = 2, 3, \dots$ by a diagram connecting the nodes 
$(M, K+1)$, $(M, K-1)$, $(M-1, K)$, $(M+1, K)$ on a $\mathbb{N} \times \mathbb{Z}$ grid, 
see Figure \ref{grid}.  
For $M=1$, we can employ a different symbol to signal the contribution of the $y|\pm$ functions, as is done by
 using a black square in Figure \ref{grid2}.
When iterating (\ref{eq:Y2}), we obtain more complex graphs such as the ones in Figures \ref{llow} and \ref{hhigh}. 
The rule to associate an equation to the graph is very simple. A white circle on the node $(M, N)$ with $M \in \mathbb{N}^+$ gives a term 
$ + \La_{M|w}( u + i N/h )$ for every horizontal link and a term $-\ln Y_{M|w}( u + i N/h ) $ for every vertical link departing from it. 
A black square on the node $(0, N)$ represents the term $L_{y|-}( u + i N/h ) - L_{y|+}( u + i N/h )$.  Figures \ref{llow} and \ref{hhigh} 
then translate into the equation:
\begin{align}
\begin{split}
\ln Y_{w|2}(u & + i(2+2N)/h)=D^{w|2}_{ +(2+2N)}(u)+\La_{w|N+3}(u + i(1+N)/h)\\
&+  L_{w|N+2}(u + i N/h)-\ln Y_{w|N+1}(u + i(N-1)/h),
\end{split}
\label{eq:def2}
\end{align}
for $N=2$ and $N=3$ respectively, where the function $D^{w|2}_{ (2+2N)}(u)$ is defined as
\bea\label{exp2}
D^{w|2}_{ (2+2N)}(u) &=  L_{y|-}(u + i2N/h)-L_{y|+}(u + i2N/h)+   L_{w|1}(u + i(2N+1)/h)\nn \\
&+\sum_{k = 1}^{N}{\Big ( } L_{w|k}(u + i(2N-k)/h)+ 2 L_{w|k+1}(u + i(2N-k+1)/h) \nn\\
&+L_{w|k+2}(u + i(2N-k + 2)/h){\Big )}.
\eea
Notice that $D^{w|2}_{ (2+2N)}(u)$ sums up the contribution of all the nodes that lie on or above the diagonal. Using the fact that the 
branch points of $\ln Y_{w|M}(u)$ closest to the real axis have $\IIm(u) = \pm M/h$, (\ref{eq:def2}) implies that
\begin{align}\label{eq:der}
\Big[ \ln{ Y }_{w|2}(u) \Big]_{ ( 2+ 2 N )} =
\left[D^{w|2}_{(2 +2N)}(u) \right]_{0} -\delta_{N, 0} \left[ \ln Y_{1|w}(u )\right]_{-1}.
\end{align}
Equations (\ref{exp2}) and (\ref{eq:der}) constitute an example of a discontinuity relation derived using only the standard Y-system. 
Notice that the term $ \left[ \ln Y_{1|w}(u) \right]_{-1} $ in the last line can 
be determined using the fundamental discontinuity relation (\ref{d1}).

As a last comment on the structure of these relations, notice that the graph in Figure \ref{llow} contains all the 
nodes present in Figure \ref{hhigh}, translated by two units upwards. This property is reflected by the following identity: 
\begin{align}
\begin{split}
&D^{w|2}_{(2N+2)}(u)-D^{w|2}_{2N}(u + i2/h)\\
&= 2 L_{w| N+1 }(u + i (N+1)/h)+L_{w|N}(u + iN/h) + L_{w|N+2}(u +  i(N+2)/h),
\end{split}
\label{eq:rulerulew}
\end{align}
for $N = 0, 1, \dots$, that allows to define the $D^{w|2}_{(2 +2 N)}(u)$ functions recursively, starting from $D^{w|2}_{2}(u)$. Notice that the 
terms on the rhs of (\ref{eq:rulerulew}) correspond to the nodes lying on the diagonal and framed by a blue rectangle in Figure \ref{hhigh}.\\
The recursive presentation given above is very convenient for the purpose of deriving the TBA from the extended Y-system, which can be done
 along the lines of \cite{Extended}. Although we do not repeat here this calculation, we list all the useful identities in Appendix \ref{list}.
\resection{ Conclusions }\label{conclusions}
In this paper we have derived, starting from the Thermodynamic Bethe Ansatz equations describing the $\Ad$ spectrum, a set of functional relations
 that characterise the analytic structure of the Y functions. Extremely similar results have been previously derived for the $\Bd$ TBA, and 
with essentially the same proof as the one contained in \cite{Extended}, one could show that the Y-system \YYs, extended by the new relations 
\DDs, is equivalent to the TBA. The advantage is that the new functional relations, contrary to the integral TBA equations, take the same form
 for all the excited states of the theory.

We expect this result to have the same applications as in the $\Bd$ case.
In particular, in that context the discontinuity relations have been used to derive rigorously excited state TBA equations \cite{Balog:disco} and 
also to prove the equivalence between the TBA and a 
much handier finite set of nonlinear integral equations of Destri-deVega type, the FiNLIE proposed in \cite{Gromov:Solving} 
(alternative simplified NLIEs have also been derived in \cite{Balog:Hybrid}). The study of the properties of the T-system presented in Appendix \ref{appT} is a preliminary step in this direction. 

Very recently a new and much simpler formulation of the $\Bd$ spectral problem has appeared, the 
$\mathbb{P}\mu$-system \cite{Pmu}. We believe that our results will be useful in achieving a similar simplification also for the theory considered here.

\medskip
\noindent{\bf Acknowledgments --}
We thank Dmytro Volin and Fedor Levkovich-Maslyuk for fruitful discussions. 
This project was  partially supported by INFN grants IS  FI11, P14, PI11,
the Italian
MIUR-PRIN contract 2009KHZKRX-007 {\it ``Symmetries of the Universe and of
the Fundamental Interactions''},
the UniTo-SanPaolo research  grant Nr TO-Call3-2012-0088 {\it ``Modern
Applications of String Theory'' (MAST)},
the ESF Network {\it ``Holographic methods for strongly coupled systems''
(HoloGrav)} (09-RNP-092 (PESC))
 and MPNS--COST Action MP1210 {\it ``The String Theory
Universe''}.
AC is supported by a City University Research Fellowship.
\appendix
\resection{ The kernels }\label{kernels}
The kernels appearing in the TBA equations \TTBA are defined as:
$$
\phi_{A, B}( u , z  ) = \frac{1}{2 \pi i } \frac{d}{d u} \ln S_{A, B}( u , z ).
$$
For different values of the indices $A$, $B$, the S-matrix elements $S_{A, B}(u, z)$ are defined as
\bea
S_{(v|M), \CQ}(u,z)&=&S_{\CQ,(v|M)}(z,u)= \left(
\frac{x(z-\frac{i}{h}\CQ)-x(u+\frac{i}{h}M )}{x(z+\frac{i}{h}\CQ)-x(u+ \frac{i}{h} M )}\right)
\left(\frac{x(z+\frac{i}{h}\CQ)}{x(z-\frac{i}{h}\CQ)} \right) \nn\\
&\times& \left(\frac{x(z-\frac{i}{h}\CQ)-x(u-\frac{i}{h} M)}{x(z+\frac{i}{h}\CQ)-x(u-\frac{i}{h}M )}\right)\prod_{j=1}^{M-1} 
\left(\frac{z-u-\frac{i}{h}(\CQ-M+2j)}{z-u+\frac{i}{h}(\CQ-M+2j)}\right)~,\\
S_M(u) &=&  \left(\frac{u-\frac{i}{h} M }{u+\frac{i}{h} M }
\right),
\eea
\eq
S_{K,M}(u)= \left( {u - \fract{i}{h} |K-M| \over u +\frac{i}{h} |K-M|} \right) \left(
 {u - \frac{i}{h} (K+M) \over u +\frac{i}{h}(K+M)} \right)
\prod_{k=1}^{\text{min}(K,M)-1} \left( {u - \frac{i}{h} (|K-M|+2k) \over u +\frac{i}{h}(|K-M|+2k)} \right)^2.
\en
Moreover, we have
\eq
S_{(y|\mp), \CQ}(u,z)=S_{\CQ, (y|\mp)}(z,u) =  \left(\frac{x(z-\frac{i}{h} \CQ)-(x(u))^{\pm 1}}{x(z+\frac{i}{h}\CQ)-(x(u))^{\pm 1}}\right) 
\sqrt{\frac{x(z+\frac{i}{h}\CQ)}{x(z-\frac{i}{h}\CQ)}}~.
\en
Notice that we have the important property 
\bea\label{important}
S_{(y| -), \CQ}( z , u_* ) = S_{(y| +), \CQ}( z , u ). 
\eea
The  elements $S_{(\CQ|\alpha),(\CQ'|\beta)}(u ,z)$ are:
\eq
S_{(\CQ|\alpha),(\CQ'|\beta)}(u,z)= S^0_{(\CQ|\alpha),(\CQ'|\beta)}(u-z) (\Sigma^{\CQ,\CQ'}(u,z))^{-1}~,
\en
where  $\Sigma^{\CQ,\CQ}$ is the improved dressing factor defined in \cite{Arutyunov:Dressing}
\eq
\Sigma^{\CQ,\CQ'}(u,z)=\prod_{k=1}^{\CQ}\prod_{l=1}^{\CQ'}\left(\frac{1-\frac{1}{x(u+\frac{i}{h}(\CQ+2-2k))x(z+\frac{i}{h}(\CQ'-2l))}}{1-\frac{1}{x(u+\frac{i}{h}(\CQ-2k))
x(z+\frac{i}{h}(\CQ'+2-2l))}}\right)\sigma^{\CQ, \CQ'}(u,z)~,
\en
and $ \sigma^{\CQ, \CQ'}(u,z) $ is the dressing factor for the direct theory\cite{Janik:DF,BHL,Volin:Review, Volin:Crossing,DHM} with both 
arguments continued to mirror kinematics. The following equivalent expression was derived in \cite{GKVI}:
\bea
\label{dress}
K_{\CQ'\CQ}^{\Sigma}(u,v)  = \frac{1}{2 \pi i}\frac{d}{du}{\ln\Sigma}_{\CQ' \CQ}(u, v)=\oint_{\bgammax} d\z \; \phi_{\CQ', y}(u, \z)
\oint_{\bgammax} d\zi \; K_{\Gamma}^{\left[2\right]}(\z-\zi)\phi_{y, \CQ}(\zi, v).~~~~~~
\eea
The definition of $S^0_{(\CQ|\alpha),( \CQ'|\beta)}(u)$ can be found in Appendix A of \cite{AdS4}, and we omit it here as it is rather lenghty 
and does not enter the computations presented in this paper. 

Finally, the following kernels appear in the computations of Section \ref{dir}:
\eq
K( u , z ) = \frac{  \sqrt{4-z^2}}{2 \pi i \; \sqrt{4- u^2}( u- z)} 
\en
satisfying 
\eq
K(z-i \CQ/ h, u) -K(z+i \CQ/h, u) = \phi_{\CQ,(y|-)}(z,u)- \phi_{\CQ,(y|+)}(z, u),
\label{K1}
\en
and
\eq
K_{\Gamma}^{[2]}(z) = {1 \over 2 \pi i } {d \over dz} \ln {\Gamma(1-ihz/2)  \over \Gamma(1+ihz/2) }.
\en
\resection{ Additional relations for the fermionic Y functions }\label{additional}
In this Appendix, we show how to derive the following identity: 
\eq
\label{eq:newdisc}
\left[\ln \left(\left( Y_{y|-}\right)\left( Y_{y|+} \right)\right) \right]_{ \pm 2M }=2\sum_{j = 1}^{M}
\left[ L_{v|j}-L_{w|j}\right]_{\pm(2M-j)}-\sum_{\CQ=1}^M \left[ L_{\CQ|I} +  L_{\CQ|II} \right]_{\pm (2M-\CQ)}, 
\en
$M = 0, 1, \dots$ from the extended Y-system \YYs, \DDs.\\
A similar calculation was presented in Appendix F of \cite{Extended} in the $\Bd$ case, but the demonstration contained a logical gap.
The amended derivation we present here can be easily adapted to the $\Bd$ case\footnote{ This answers the question raised in the footnote (43) 
of \cite{Gromov:Solving}, and shows that (\ref{eq:newdisc}), and its analogue F.5 of \cite{Extended}, can be derived from the extended Y-system 
and do not need to be independently postulated.}.

We start by considering the following relation, which is an immediate consequence of the Y-system equation (\ref{Yysystem}) and is valid for any 
$N\in\mathbb{Z}$:
\begin{align}\label{Ymp}
&~~~2\left[ \ln\left(Y_{y|-}\right) \right]_{ \pm 2N } + 2\left[ \ln\left( Y_{y|-} \right) \right]_{ \pm (2 N  - 2 )} \nn\\
&~~~=\left[ \ln\left( \left(Y_{y|-}\right) \left( Y_{y|+}  \right) \right) + \ln\left( Y_{y|-}/ Y_{y|+}\right)\right]_{ \pm 2N } + 
\left[ \ln\left( \left( Y_{y|-} \right)\left(Y_{y|+} \right)\right) + \ln\left(Y_{y|-} / Y_{y|+}\right) \right]_{ \pm ( 2 N - 2 ) }\nn\\
&~~~ = 2\left[ \La_{v|1}-\La_{w|1} - \sum_{\alpha = I, II} L_{1|\alpha}\right]_{ \pm ( 2 N - 1) } = 2\left[ L_{v|1}-L_{w|1} + 
\ln\left( { Y_{v|1} /Y_{w|1} }\right)- \sum_{\alpha = I, II} L_{1|\alpha } \right]_{ \pm ( 2 N - 1 ) },
\end{align}
where we have set $\La_a(u)=\ln(1+Y_a(u))$. Next, we use the following identity, which can be derived iterating the Y-system equations 
(\ref{yw}) and (\ref{yv}) (for example, one can use the graphical method described in Section \ref{graphical}):
\begin{align}
\begin{split}\label{numb}
\left[\ln\left({Y_{v|1} /Y_{w|1}}\right)\right]_{\pm (2N-1)}&=\left[\ln\left(Y_{y|-} / Y_{y|+}\right)\right]_{\pm(2N-2)}-\sum_{\alpha = I, II} 
\sum_{\CQ =2}^{N}\left[L_{\CQ | \alpha }\right]_{\pm (2N-\CQ)}\\
+\sum_{j=1}^{N-1}\sum_{k \geq 1 }&\I_{j k}\left[L_{v|j}-L_{w|j}\right]_{\pm(2N-1-k)}+\left[L_{v|N}-L_{w|N}\right]_{\pm N} - \delta_{N, 1} 
\left[ \ln\left({Y_{v|1} /Y_{w|1}}\right)\right]_{\mp 1},
\end{split}
\end{align}
for $N=1, 2, \dots$
and where $A_{j k} = \delta_{j, k+1} + \delta_{j, k-1}$.
Combining (\ref{numb}) and (\ref{Ymp}) we find:
\begin{align}\label{alas}
\left[\ln \left({Y_{y|-} Y_{y|+}}\right)\right]_{\pm 2N}+&\left[\ln\left({Y_{ y|-} Y_{y|+}}\right)\right]_{\pm ( 2N-2 ) }\nn\\
=&\; 2\sum_{j=1}^{N}\left[L_{v|j}-L_{w|j}\right]_{\pm(2N-j)}+2\sum_{j=1}^{N-1}\left[L_{v|j}-L_{w|j}\right]_{\pm(2N-2-j)}\nn\\
&-2 \sum_{\alpha = I, II} \sum_{\CQ =1}^{N}\left[L_{\CQ|\alpha} \right]_{\pm (2N-\CQ)}- \left[ \ln\left( Y_{y|-}/ Y_{y|+}\right)\right]_{ \pm 2N } 
\nn\\
&+ \left[ \ln\left( Y_{y|-}/ Y_{y|+}\right)\right]_{ \pm ( 2N-2) } - 2\;\delta_{N, 1} \left[ \ln\left({Y_{v|1} /Y_{w|1}}\right)\right]_{\mp 1}.
\end{align}
Now we can invoke three of the fundamental relations, namely the two identities in (\ref{d3}) (leading to 
$\left[ \ln\left({Y_{v|1} /Y_{w|1}}\right)\right]_{\mp 1} = \ln{ Y_{y|-} / Y_{y|+} }$) and (\ref{d2}), to evaluate the terms in the 
last two lines of (\ref{alas}). The result is
\begin{align}
\label{recur}
\left[\ln \left({Y_{y|-} Y_{y|+}}\right)\right]_{\pm 2N}+&\left[\ln\left({Y_{ y|-} Y_{y|+}}\right)\right]_{\pm ( 2N-2 ) }\\
=&\; 2\sum_{j=1}^{N}\left[L_{v|j}-L_{w|j}\right]_{\pm(2N-j)}+2\sum_{j=1}^{N-1}\left[L_{v|j}-L_{w|j}\right]_{\pm(2N-2-j)}\nn\\
&-\sum_{\alpha = I, II} \left( \sum_{\CQ =1}^{N}\left[L_{\CQ|\alpha } \right]_{\pm (2N-\CQ)} + 
 \sum_{\CQ =1}^{N - 1}\left[L_{\CQ|\alpha} \right]_{\pm (2N-\CQ - 2)} \right) ,\hspace{0.3cm} N = 0, 1, \dots \nn
\end{align}
Finally, because  of (\ref{d3}) we have $\left[\ln \left({Y_{y|-} Y_{y|+}}\right)\right]_{ 0 } = 0$. Using this fact we can solve (\ref{recur}) recursively and finally get (\ref{Ymp}).
\resection{ A list of useful identities }\label{list}
For the sake of completeness, below we list a number of identities that can be derived using only the structure of the basic Y-system \YYs. As shown 
in \cite{Extended}, these relations are useful for the purpose of rederiving the TBA equations.

The discontinuities of the $ \ln Y_{v|M}(u)$ functions satisfy:
\begin{align}
\Big[ \ln{ Y }_{w|M}(u) \Big]_{ \pm( M + 2 N)} = 
\left[D^{w|M}_{\pm(M +2 N)}(u)-\delta_{N, 0} \ln Y_{1|w}(u \mp  i/h)\right]_{0}
\end{align}
for $N = 0, 1, \dots $, where the $D^{w|M}_{\pm(M + 2 N) }(u)$ functions are defined iteratively by
\bea \label{eq:rulew}
&&D^{w|M}_{\pm(2 N+M)}(u)-D^{w|M}_{\pm(2 N +M-2)}(u\pm i2/h)\\
&&= 2\sum_{k= N+1}^{M+ N-1} L_{w|k}(u \pm i k/h)+L_{w| N}(u\pm i N/h) + L_{w|M+N}(u\pm i(M+N)/h),\nn
\eea
with
\bea
D^{w|M}_{\pm M}(u)= L_{y|-}(u)-L_{y|+}(u) +\sum_{ k =1}^{M-1} L_{w| k}(u \pm i k/h).
\label{eq:firstwfunction}
\eea
The discontinuities of the $ \ln Y_{v|M}(u)$ functions satisfy
\bea
\left[ \ln Y_{v|M}(u) \right]_{\pm ( M + 2 N )}&=&\left[D^{v|M}_{\pm ( M + 2 N )}(u)- \delta_{N, 0} \; \ln Y_{v|1}( u \mp i/h ) \right]_{0}, 
\eea
and the $ D^{v|M}_{\pm(M + 2 N )}(u) $ functions are defined by
\begin{align}\label{eq:rulev} 
D^{v|M}_{\pm(M + 2 N )}(u)&-D^{v|M}_{\pm(M + 2 N -2)}(u\pm i2/h) = L_{v|N}(u\pm i N/h)+L_{v|M+N}(u\pm i(M+N)/h) \nn\\ &+2 
\sum_{k=N+1}^{M+ N-1}L_{v|k}(u \pm ik/h)- \sum_{\CQ= N+1}^{M+N} \; \sum_{\alpha = I, II} \; L_{\CQ | \alpha }(u \pm i\CQ/h),
\end{align}
with
\bea
D^{v|M}_{\pm M}(u) = \La_{y|-}(u)
-\La_{y|+}(u)+ \sum_{ k = 1 }^{M-1} L_{v| k }(u \pm i k/h),
\eea
The discontinuities of the $ \ln Y_{\CQ| \alpha}$ functions satisfy
\bea
&&\left[ \ln Y_{ \CQ | \alpha }(u)\right]_{ \pm ( \CQ + 2 \CJ )  } = \left[ D^{ \CQ | \alpha }_{ \pm\CQ }(u) -  \delta_{ \CJ , 1}
 \ln Y_{1|\beta}(u \mp i / h )\right]_{0} 
\label{eq:otherD}
\eea
for $\CJ =1,2,\dots$, where $\alpha \in \left\{ I, II \right\}$, $\beta \in \left\{ I, II \right\}$, 
$\alpha = \beta $ if $\CQ$ is odd and $\alpha \neq \beta$ if $\CQ $ is even. 

The $ D^{\CQ|\alpha}_{\pm( \CQ + 2 \CJ )}(u)$ functions are defined by\footnote{ Given a real number $r \in \mathbb{R}$, we denote its integer
 part as $\lfloor r \rfloor \in \mathbb{Z}$. }
\begin{align}
\begin{split}
D^{\CQ | \alpha }_{\pm(\CQ+2 \CJ)}(u)&-D^{\CQ | \alpha }_{\pm(\CQ+2 \CJ -2)}(u\pm
i2/h)= L_{\CJ | \beta }(u\pm i\CJ/ h ) +L_{\CQ+\CJ | \alpha }(u\pm
i(\CQ+\CJ)/ h )\\
&+ 2 \; \sum_{ k = 1 }^{\lfloor \frac{\CQ - 1}{2} \rfloor } L_{\CQ + \CJ - 2 k | \alpha }( u \pm i ( \CQ + \CJ  - 2 k  )/ h) - 
\sum_{ N =\CJ}^{\CQ + \CJ-1}L_{v|N}(u\pm iN/ h )\\
&+ 2 \; \sum_{ k = 1 }^{\lfloor \frac{\CQ}{2} \rfloor } L_{ \CQ + \CJ +1 -  2 k  | \gamma }( u \pm i ( \CQ + \CJ + 1 - 2 k )/ h) , 
\end{split}
\end{align}
where $\alpha, \beta, \gamma \in \left\{ I, II \right\}$, $\gamma \neq \alpha$, $\alpha = \beta $ if $\CQ$ is odd and $\alpha \neq \beta$ 
if $\CQ $ is even, and:
\bea\label{eq:firstQ}
D^{\CQ | \alpha }_{\pm \CQ}(u) &=&\sum_{ k = 1 }^{\lfloor \frac{ \CQ - 1 }{2} \rfloor } L_{\CQ - 2 k | \alpha }( u \pm i(  \CQ  - 2 k  )/h) + 
\sum_{ k = 1}^{\lfloor \frac{ \CQ }{2} \rfloor  } L_{\CQ + 1 - 2 k | \gamma }( u \pm i(  \CQ + 1 - 2 k  )/h) \nn\\
&&-\left(\sum_{N = 1}^{\CQ-1}L_{v|N}(u \pm iN/ h ) + L_{y}(u)\right),
\eea
with $\alpha, \gamma \in \left\{ I, II \right\}$ and $\gamma \neq \alpha$.
\setcounter{footnote}{0}
\resection{The T-system}\label{appT}
In this section we will reconsider the discontinuity relations \DDs from the point of view of the T-system, as done in 
\cite{Balog:disco,Gromov:Solving} in the case of $\Bd$. In particular, following \cite{Gromov:Solving}, we show how to encode the analytic 
content of the TBA equations into a set of very symmetric constraints for the T functions.\\
The Y-system of $\Ad$ is naturally related to the diagram represented in Figure \ref{figY}. In fact, let us associate a Y function to every 
node of the diagram, using an additional index $\alpha \in \left\{ I, II\right\}$ to distinguish the functions living on the two wings. 
Then the Y-system relations \YYs {} can be written in a universal form using the incidence matrix of the diagram:\footnote{ 
However, we point out that the diagram in Figure \ref{figY} does not capture the crossing between the two wings in the lhs of the Y-system equations for the nodes $(n , 0, \alpha)$, $\alpha = I, II$. For non-symmetric states such that $ Y_{n|I} \neq Y_{n|II}$, we need to keep track of this important subtlety.
}
\bea
Y_{a, s}( u + i/h ) Y_{a, s}(u - i/h ) &=& \frac{ ( 1 + Y_{a, s+1}( u ) )( 1 + Y_{a, s-1}( u ) )}{( 1 + 1/Y_{a+1, s}( u ) )
(1 + 1/Y_{a-1, s}( u ) )},\hspace{1cm}
 s > 1, (a,s) \neq(2,2)\nn\\
Y_{a, 1}( u + i/h ) Y_{a, 1}(u - i/h ) &=& \frac{ ( 1 + Y_{a, 2}( u ) )( 1 + Y_{a, 0}^{I}( u ) )
( 1 + Y_{a, 0}^{II}( u ) )}{( 1 + 1/Y_{a+1, 1}( u ) )(1 + 1/Y_{a-1, 1}( u ) )},\nn\\
Y_{a, 0}^{\alpha}( u + i/h ) Y_{a, 0}^{\beta}(u - i/h ) &=& \frac{ ( 1 + Y_{a, 1}( u ) )}{( 1 + 1/Y_{a+1, 0}^{\alpha}( u ) )
(1 + 1/Y_{a-1, 0}^{\beta}( u ) )}.
\eea
Notice that we have to exclude the node $(2,2)$, as there is no local Y-system equation in this case. The Y functions in the double index 
notation are related to the ones used in the rest of this paper and in \cite{AdS4} by: 
\bea
Y_{\CQ , 0}^{\alpha} &=& 1/Y_{\CQ | \alpha}, \hspace{1cm}\text{ for }\CQ \geq 1 , \alpha = I, II \nn,\\
Y_{1 , 1} &=& 1/Y_{ y | - }, \nn\\
Y_{2 , 2} &=& Y_{ y | + }, \nn\\
Y_{n , 1} &=& 1/Y_{ v | n-1 }, \hspace{0.7cm}\text{ for } n \geq 2 \nn,\\
Y_{1 , n} &=& Y_{ w | n-1 }, \hspace{1cm} \text{ for } n \geq 2 .\nn
\eea
The T functions live on a lattice obtained by adding extra nodes to the diagram in Figure \ref{figY}. We denote them as
\bea
T_{n, s} ,        &\text{ with }(n , s) \in \mathbb{N} \times \mathbb{N}^+ ,\hspace{0.5cm} s \leq 2 \text{ or } n \leq 2 ,\nn\\
T_{n, l}^{\alpha},&\text{ with }(n , l) \in \mathbb{N} \times \left\{ -1 , 0 \right\} , \hspace{0.4cm} \alpha \in \left\{I, II \right\},\nn
\eea
and they are assumed to be zero when the indices are outside the domain indicated above.
 They are related to the Y functions by:
\bea
\label{eq:Ydef}
Y_{a, s}(u) &=& \frac{ T_{a, s+1}(u) T_{a, s-1}(u) }{ T_{a+1, s}(u) T_{a-1, s}(u) },\hspace{1cm}\text{ for }s\geq 2, a\geq 1, \\
Y_{a, 1}(u) &=& \frac{ T_{a, 2}(u) T_{a, 0}^{I}(u)T_{a, 0}^{II}(u) }{ T_{a+1, 1}(u) T_{a-1, 1}(u) },\hspace{1cm}\text{ for } a\geq 1,\nn\\
Y_{a, 0}^{\alpha}(u) &=& \frac{ T_{a, 1}(u) T_{a, -1}^{\beta}(u) }{ T_{a+1, 0}^{\alpha}(u) T_{a-1, 0}^{\beta}(u) },\hspace{1cm}\text{ for } a\geq 1 , \hspace{0.3cm} \alpha, \beta \in \left\{ I, II \right \}, \beta \neq \alpha. \nn
\eea
The Y-system is satisfied provided the T functions obey discrete Hirota equations on the lattice: the T-system. For the $\Ad$ 
diagram, the T-system relations take the usual form for $s \geq 2 $:
\bea
\label{T1}
T_{n, s}( u + i/h ) T_{n, s}(u - i/h ) = ( 1 - \delta_{n, 0} )T_{n + 1 , s }(u) T_{n-1, s}(u) + T_{n, s-1}(u) T_{n, s+1}(u),
\eea
while there is a cubic term when $s = 1$
\bea\label{T2}
T_{n, 1}( u + i/h ) T_{n, 1}(u - i/h ) = ( 1 - \delta_{n, 0} )T_{n+1 , 1 }(u) T_{n-1, 1}(u) + T_{n , 0}^{I}(u) T_{n, 0}^{II}(u) T_{n, 2}(u), 
\eea
and the equations with $s = -1 , 0$ are
\bea\label{T3}
T_{n, 0}^{\alpha}( u + i/h ) T_{n, 0}^{\beta}(u - i/h ) &=& ( 1 - \delta_{n, 0} )T_{n+1 , 0 }^{\alpha}(u) T_{n-1, 0}^{\beta}(u) + T_{n , -1}^{\beta}(u) T_{n, +1}(u),\nn\\
T_{n, -1 }^{\alpha}( u + i/h ) T_{n, -1}^{\beta}(u - i/h ) &=& T_{n+1 , -1 }^{\alpha}(u) T_{n-1, -1}^{\beta}(u) \hspace{0.3cm} (n \neq 0)\;,
\eea
with $\alpha, \beta \in \left\{ I, II \right \}$ and $\beta \neq \alpha $. 

It is well known that the same solution to the Y-system is parametrised by a large family of equivalent solutions to the T-system, connected 
by gauge transformations. A generic gauge transformation preserving the validity of the T-system (\ref{T1}-\ref{T3}) and leaving invariant 
the Y functions can be written as follows:
\bea
\label{eq:gaugetransf}
&&T_{n, 0}^{\alpha} \rightarrow f_{\alpha}^{\left[  n \right] } \; g_{\alpha}^{\left[  n \right] }  \; h^{\left[  -n \right] } \; j^{\left[  -n \right] }\; 
T_{n, 0}^{\alpha},\hspace{3cm}\text{  for  }\alpha \in \left\{ I, II \right\}, n \in\mathbb{ N },\\
&&T_{n, - 1}^{\alpha} \rightarrow \frac{(f_{\alpha})^{\left[  n-1 \right] } }{ (f_{\alpha})^{\left[  n+1 \right] }  } 
\frac{(g_{\alpha})^{\left[  n+1 \right] } }{ (g_{\alpha})^{\left[  n-1 \right] }  } \frac{(h)^{\left[  -n-1 \right] } }{ (h)^{\left[  -n+1 \right] }}
 \frac{(j)^{\left[  -n+1 \right] } }{ (j)^{\left[  -n-1 \right] }  } \;T_{n, -1}^{\alpha},\hspace{0cm}\text{  for  }\alpha, \beta \in 
\left\{ I, II \right\}, \alpha \neq \beta \text{ and } n \in\mathbb{ N } ,\nn\\
&&T_{n, s} \rightarrow ( f_{I} \; f_{II} )^{\left[  n + s \right]} \; ( g_{I} \; g_{II} )^{\left[  n - s \right]} \; 
( h^2 )^{\left[  -n + s \right]} \; ( j^2 )^{\left[  -n - s \right]} \; T_{n, s},\hspace{1cm}\text{ for }s \in \mathbb{ N }^+ , n\in\mathbb{ N },
\nn 
\eea
where $f_{I}, f_{II}, g_{I}, g_{II} , h, j$ are arbitrary functions and we have adopted the notation: 
$ A^{\left[ a \right]}( u ) \equiv A( u + i a/h )$ to denote imaginary shifts in the rapidity.\\

Let us now translate the discontinuity relations in terms of the T functions. 
A straightforward calculation shows that the two equations in (\ref{d3}) can be rewritten in the following form:
\bea
\label{eq:wDis}
\frac{ T_{1, 1}( u - i/h ) T_{1, 1}( u_* + i/h ) }{ T_{2, 2}( u - i/h ) T_{2, 2}( u_* + i/h ) }\frac{T_{2, 3}(u)}{ T_{0, 1}(u) } = 
\frac{ E( u )}{E( u_* ) }
\eea
and
\bea
\label{eq:vDis}
\frac{ T_{2, 2}( u - i/h ) T_{2, 2}( u_* + i/h ) }{ T_{1, 1}( u - i/h ) T_{1, 1}( u_* + i/h ) }\frac{ T_{1, 0}^{I}(u) T_{1, 0}^{II}(u) }{T_{3, 2}(u)}
 = \frac{ F( u )}{F( u_* ) },
\eea
where we have defined 
\bea
E(u) &=& T_{1, 3}( u + i/h ) T_{0, 2}( u_* + i/h ), \nn\\
F(u) &=& T_{3, 1}( u_* + i/h )  \prod_{\alpha= I, II}T_{2, 0}^{\alpha}( u + i/h ).
\eea
It was shown in \cite{Balog:disco,Gromov:Solving} in the $\Bd$ case that the remaining, infinitely many discontinuity relations can be greatly simplified by making appropriate assumptions on the analyticity strips of the T functions. 
In particular, in \cite{Gromov:Solving} it was shown that there exist two special gauges where these constraints take a very symmetric form. 
They were denoted with the fonts $\bfT$ and $\bbT$, with the $\bfT_{n, s}$ functions having particularly convenient analytic properties in the 
upper band of the diagram defined by  $ n \geq s$ and the $\bbT_{n, s}$ functions being particularly well behaved in the right band defined 
by $ s \geq n$. We conclude this section by showing how the same can be achieved in the present case.\\ 

We follow very closely Appendix C of \cite{Gromov:Solving}, where very similar calculations are presented. Let us borrow a useful notation: we denote 
as $\mathcal{ A }_n$ the class of functions meromorphic in the strip $| \IIm(u) |< n/h $. In general, we expect T functions in $\mathcal{A}_n$ to 
have branch points at $\pm 2 + in/h$, $\pm2 -in/h$. Then, we start by considering a gauge, which we denote generically with the font $T \equiv t$, such that the $t$ functions are 
real\footnote{ In the case of non-symmetric states such that $Y_{n, 0}^{I} \neq Y_{n, 0}^{II}$, the requirement of reality has 
to be replaced with $\bar{t}^{I}_{n ,  0} = t^{II}_{n, 0}$. } and satisfy
\bea\label{require}t_{n, -1}^{\alpha} = 1,\hspace{1cm}t_{n, 0}^{\alpha} \in \mathcal{ A }_{n+1},\hspace{1cm}t_{n, 1} \in \mathcal{ A }_{n},\eea
for $n \in \mathbb{N}, \alpha \in \left\{ I, II \right\}$. \\
Let us now consider the discontinuity relation (\ref{d2}). When expanding the factors $(1 + 1/Y_{v|n})$ in terms of the $t$ functions, their jump discontinuities cancel 
pairwise (see similar calculations in \cite{Balog:disco,Gromov:Solving}) and we find that this condition is equivalent 
to\footnote{ To be precise, the requirement $t_{n, 0}^{\alpha} \in \mathcal{ A }_{n+1}$ is enough to prove (\ref{eq:B0}). The other conditions
 in (\ref{require}) have been added for future convenience. } 
\bea
\label{eq:B0}
\left[ \; \ln\textbf{b}(u) \; \right]_{2 N } = 0,\hspace{1cm}N=1,2,\dots .
\eea
where 
\bea\label{eq:B}
{\bf b }(u) =  Y_{1, 1}(u) Y_{2, 2}(u) \prod_{\alpha = I, II}\frac{ t_{0, 0}^{\alpha}(u - i/h ) }{ t_{1, 0}^{\alpha}( u ) } =  
\frac{ t_{2, 3}( u ) }{ t_{3, 2}( u ) }\frac{ t_{0, 0}^{I}(u - i/h ) \; t_{0, 0}^{II}(u - i/h ) }{ t_{0, 1}( u ) }.
\eea
Therefore the function $ {\bf b }(u) $ defined above is meromorphic in the upper half plane. Notice that $ {\bf b }(u) $ is still a 
gauge-dependent quantity. However, following \cite{Gromov:Solving}, let us define a transition function $f$, analytic for $\IIm(u) > -1/h$, 
such that
\bea
\label{eq:deff}
{\bf b }(u) \equiv \frac{ f^2( u + i/h )}{ f^2( u - i/h )}.
\eea
Then, making a gauge transformation of the form\footnote{ Notice that we denote with $\bar{f}$ the complex conjugate function such
 that $\bar{f}(u) = (f(u^*))^*$. }
\bea
\label{eq:gaugeboldT}
&&{ \bf T }_{n, s}  =  f^{\left[  n + s \right]} \;  f ^{\left[  n - s \right]} \;  \bar{ f }^{\left[  -n + s \right]} \;  
\bar{ f }^{\left[  -n - s \right]} \; t_{n, s},\hspace{1cm}\text{ for }s \in \mathbb{ N }^+ , n\in\mathbb{ N } , \nn\\
&&{ \bf T }_{n, 0}^{\alpha} = f^{\left[  n \right] } \; \bar{ f }^{\left[  -n \right] }\;
 t_{n, 0}^{\alpha},\hspace{3.5cm}\text{  for  }\alpha \in \left\{ I, II \right\}, n \in\mathbb{ N },\nn \\
&&{ \bf T }_{n, - 1}^{\alpha} = t_{a, -1}^{\alpha} = 1,\hspace{4cm}\text{  for  }\alpha \in \left\{ I, II \right\},  n \in\mathbb{ N } ,
\eea
we find a gauge satisfying 
\bea\label{eq:one} 
\frac{ {\bf T }_{2, 3}( u ) }{ {\bf T }_{3, 2}( u ) }\frac{ {\bf T }_{0, 0}^{I}(u - i/h ) \;
 {\bf T }_{0, 0}^{II}(u - i/h ) }{ {\bf T }_{0, 1}( u ) } = 1 .
\eea 
Following \cite{Gromov:Solving}, let us show how to deduce the following very special properties of the $\bfT$ gauge:
\begin{enumerate}
\item The $\bf{T}$ functions are real, and we have 
\bea\label{eq:bbTbasic}
{ \bf T }_{n, -1}^{\alpha} = 1 \hspace{0.5cm}  { \bf T }_{n, 0}^{\alpha} \in \mathcal{ A }_{n+1} \hspace{0.5cm} { \bf T }_{n, 1} 
\in \mathcal{ A }_{n} \hspace{0.5cm} { \bf T }_{n, 2} \in \mathcal{ A }_{n-1} ,\hspace{0.5cm}n\in \mathbb{ N }, \alpha \in \left\{ I, II \right\}.
\eea 
\item 
The two quantities $\bfT_{0, 0}^I$, $\bfT_{0, 0}^{II}$ are periodic:
\bea\label{eq:perio}
{\bf T}_{0 , 0}^{\alpha}( u + i/h ) = {\bf T}_{0 , 0}^{\alpha} ( u - i/h ) ,\hspace{0.5cm}\alpha \in\left\{I, II\right\}.
\eea 
\item Finally, the $\bf{T}$ functions enjoy the following ``group-theoretical'' properties:
\bea
{\bf T }_{0 , n } &=& ( {\bf T}_{0 , 0}^{I} {\bf T}_{0 , 0}^{II} )^{[n]}\hspace{1cm}n = 1, 2, \dots \label{eq:T0n},\\
{\bf T}_{n , 2} &=& {\bf T}_{2 , n}, \hspace{2cm}n = 2, 3, \dots \label{reflection}.
\eea 
\end{enumerate}
Here is a brief summary of the proof. Property 1) follows from the fact the transformation is real and does not change the analyticity domains
of the $t$ functions.
Then, the complex conjugate of (\ref{eq:one}) implies that the product $\bfT_{0, 0}^I \bfT_{0, 0}^{II}$ is periodic, and 
from ${\bf T}_{0 , -1}^{I} = {\bf T}_{0 , -1}^{II} = 1$ and the product of the T-system equations for the $(0, 0, I)$ and $(0, 0, II)$ 
nodes we find $\bfT_{0, 1} = ( \bfT_{0, 0}^{I} \bfT_{0, 0}^{II} )^{\left[+1\right]}$. Comparing this equation with the T-system 
at one of the above mentioned nodes, we find that not only their product, but each of the functions $\bfT_{0, 0}^I$ and $\bfT_{0, 0}^{II}$ 
is periodic, thus establishing property 2). Moreover, (\ref{eq:one}) now implies ${\bf T}_{3 , 2} = {\bf T}_{2 , 3}$. 
Equations (\ref{eq:T0n}-\ref{reflection}) for general $n$ can be demonstrated by iterating the T-system and using (\ref{eq:perio}).\\

\noindent In the rest of this Section we also make the crucial hypothesis that it is possible to choose 
\bea\label{eq:IequalsII} 
\bfT_{0, 0}^{I} = \bfT_{0, 0}^{II}.
\eea
Because many of the following results depend on this assumption, it is worth making a comment. 
Condition (\ref{eq:IequalsII}) is certainly true in the important subsector of the symmetric states such that $Y_{n|I} = Y_{n|II} $ 
$\forall n$, which includes the best-studied case of the $\mathfrak{sl}_2$ states. Moreover we argue that this choice can be made even for some non-symmetric subsectors and possibly for all states. 
We reason as follows. Even if $\bfT_{0, 0}^I \neq \bfT_{0, 0}^{II}$, the ratio $w(u) = \bfT_{0, 0}^{I} / \bfT_{0, 0}^{II}$ 
is necessarily meromorphic, because the property $\Delta^{I} = \Delta^{II}$ proved in Section \ref{sec2} 
implies that $ \ln \bfT_{0, 0}^I$ and $\ln \bfT_{0, 0}^{II}$ have the same 
discontinuities\footnote{ In fact, using the identity $ \bfT_{0, 0}^{\alpha} = \bfT_{1,1}/( Y_{1, 0}^{\beta} \bfT_{2, 0}^{ \beta } )$ 
($\alpha \neq \beta$ ) we get $ \left[ \ln \bfT_{0, 0}^{\alpha} \right]_1 = \left[ \ln \bfT_{1, 1} \right]_1 - \Delta$, 
$\alpha = I, II$, where $\Delta = \Delta^I = \Delta^{II}$. Therefore $ \left[ \ln \bfT_{0, 0}^{I} \right]_1 = 
 \left[ \ln \bfT_{0, 0}^{II} \right]_1 $ and because of the periodicity this is sufficient to prove that $\bfT_{0,0}^I/\bfT_{0, 0}^{II}$ is meromorphic. }.
Therefore, we can define a gauge transformation that sets $\bfT_{0, 0}^{I} = \bfT_{0, 0}^{II}$ by 
taking $g_{I} = \sqrt{ \bfT_{0 , 0}^I / \bfT_{0, 0}^{II} }$ and $g_{II} = 1/g_I$ in (\ref{eq:gaugetransf}), with all the other 
transition functions being unity. Notice that this transformation does not spoil any other property of the $\bfT$ gauge, on the condition 
that $g_I(u) = \sqrt{w(u)}$ is still meromorphic and no new branch cuts are introduced by the square root. We believe that this is indeed the case for the 
physical solutions to the $\Ad$ T-system. \\
{}Finally, here we leave open the problem of proving the uniqueness of the $\bfT$ gauge. However, by analogy with the $\Bd$ case it is natural to expect that it can be fixed completely
( modulo a constant rescaling of the form $\bfT_{ n , 0}^{\alpha} \rightarrow k \bfT_{n , 0}^{\alpha} $, $ \bfT_{n , s} \rightarrow k^2 \bfT_{n , s} $, with $k \in \mathbb{ R } $ ) by adding the further requirement that the ${ \bf T }$ functions do not have poles and have the minimal amount of zeroes in their analyticity strips.\\

To further underline the analogy with the $\Bd$ case treated in \cite{Gromov:Solving}, it is useful to introduce the notation 
\bea
\mathcal{F} = {\bf T}_{0 , 0}^{I} = {\bf T}_{0 , 0}^{II}.
\eea 
The absence of the square root in this definition, as compared to the $\Bd$ case, is simply due to the different 
structure of the Y-system, but, as we will see, $\mF$ plays the same r$\hat{\text{o}}$le in many respects.\\
An interesting observation is that, as already noticed in \cite{Gromov:Solving}, 
$\mathcal{F}$ is strictly related to the dressing factor. In fact, from the regularity strips of ${ \bf T}$ functions we expect 
$\mathcal{F}$ to have the closest branch points at distance $\pm i/h$ from the real axis. From (\ref{eq:T0n}) and the 
identity $Y_{1,1} Y_{2,2} = ( T_{1,0}^I T_{1,0}^{II} T_{2,3})/(T_{0,1} T_{3,2})$ it is possible to 
prove
\bea\label{eq:discoF}
 Y_{1, 1}(u) Y_{2, 2}(u) = \frac{ \mathcal{F}( u_* + i/h ) }{ \mathcal{F}( u + i/h ) }
\eea
and, because of the periodicity (\ref{eq:perio}), $\left[ \ln\mathcal{F} \right]_1 = \left[ \ln\mathcal{F} \right]_{( 1 + 2 n )} =  
-\ln{ Y_{1, 1} Y_{2, 2} }$ $\forall n \in \mathbb{Z}$. 
According to (\ref{eq:finalD4}), a periodic jump discontinuity equal to
$\pm\ln{ Y_{1, 1}Y_{2, 2}}$ characterises precisely the contribution of the dressing factor to the TBA equations.\\

Following \cite{Gromov:Solving}, let us now introduce a new gauge $\mathbb{T}$ by\footnote{ Although this transformation has a quite unusual form, it defines a new solution to the T-system thanks to the periodicity of $\mathcal{ F }$. } 
\bea\label{eq:bbtrans}
\mathbb{T}_{n , s} &=& (-1)^{n(s+1)}
{\bf T }_{n , s } \left(  \mathcal{F}^{\left[ n + s \right] } \right)^{ n-2 } ,\hspace{1cm}s\geq 1\nn\\
\mathbb{T}_{n , 0}^{\alpha} &=& (-1)^{n}
{\bf T }_{n , 0 }^{\alpha} \left( \sqrt{ \mathcal{F} ^{\left[ n\right] } } \right)^{ n-2 }\nn\\
\mathbb{T}_{n , -1}^{\alpha} &=& {\bf T }_{n , -1 }^{\alpha}  = 1. 
\eea
From (\ref{eq:bbtrans}), it follows immediately that the $\bbT$ functions are real and satisfy 
\bea\label{eq:bbprop1}
\mathbb{ T }_{0, s} = 1 \text{  for  }s \geq -1,\hspace{0.5cm}\mathbb{ T }_{2, s} = { \bf T }_{2, s} \in \mathcal{A}_s \text{  for  }s \geq 2,
\hspace{0.5cm}\mathbb{T}_{1,1} \in \mathcal{A}_1. 
\eea
Moreover, it is possible to show that
\bea\label{eq:bbprop2}
\bbT_{1, s} \in \mathcal{A}_s, \hspace{1cm}s\geq 1.
\eea
In fact, using (\ref{eq:T0n}-\ref{reflection}) and the transformation (\ref{eq:bbtrans}), one finds
\bea\label{eq:previousratio}
 \frac{Y_{11}}{Y_{22}} = \frac{T_{3,2}}{T_{2,3}} \frac{  T_{1,2}^2  T_{1,0}^{I} T_{1,0}^{II} }{ T_{0, 1} T_{2,1}^2 } =
\frac{ { \mathbb{ T } }_{1,2}^2  { \bf T}_{1,0}^{I} { \bf T}_{1,0}^{II} }{ { \bf T }_{2,1}^2 }.
\eea
Recalling the analyticity strips of the ${\bf T}$ functions and remembering that $Y_{11}/Y_{22} = 1/( Y_{y|-} Y_{y|+} ) \in \mathcal{A}_2$, 
this shows that $\mathbb{ T }_{1,2} \in \mathcal{A}_2$. The analyticity strips for $\bbT_{1, s}$, $s \geq 3$ can be established, for example, 
by considering the various identities (\ref{eq:newdisc})\footnote{ Let us exemplify the derivation by showing that $\bbT_{1, 3} \in \mathcal{A}_3$. 
The first subcase of (\ref{eq:newdisc}) can be written as 
$$
\left[ \ln \frac{Y_{11}}{Y_{22}} \right]_{\pm 2} = 2 \left[ \ln\frac{( 1 + 1/Y_{1,2} )}{( 1 + Y_{2,1} )} \right]_{\pm 1} + \sum_{\alpha}  
\left[ \ln( 1 + Y_{1, 0}^{\alpha} )\right]_{\pm 1}.
$$
Expressing $( 1 + 1/Y_{1,2} )$ in the $\bbT$ gauge and $( 1 + Y_{2,1} )$, $( 1 + Y_{1,0} )$ in the $\bfT$ gauge, the above expression becomes
\bea
\left[ \ln \frac{Y_{11}}{Y_{22}} \right]_{\pm 2} = \left[ \ln{\frac{ \bbT_{1,2}^2 \bfT_{1,0}^{I} \bfT_{1,0}^{II}  }{\bfT_{2,1}^2 } }\right]_{ \pm 2} - \left[ \ln{\bbT_{1,3}^2 }\right]_{ \pm 1 },\nn
\eea
and comparing this result with (\ref{eq:previousratio}) we deduce that the last term on the rhs vanishes. Therefore $\bbT_{1, 3}$ has no branch
 points with $\IIm(u) = \pm1/h$ and using $\bbT_{1, 2}\in\mathcal{A}_2$ the T-system implies $\bbT_{1,3} \in \mathcal{A}_3$. }, which were derived 
in Appendix \ref{additional}. \\
Now let us consider the set of discontinuity relations (\ref{d1}). Repeating the derivation of Section D.3 in \cite{Gromov:Solving}, one can show that these relations can be rewritten as
\bea
\label{eq:Deltadis}
\left[   
\Delta^{\alpha}(u) + \ln{\frac{ \bfT_{0, 1}(u) }{ \bfT_{1, 1}( u + i/h ) }}\right]_{2 N } = -\ln{  Y_{1, 1}(u) Y_{2, 2}(u) }, \hspace{1cm} N \in \mathbb{N}^+, 
\eea
and using the identity
$$
\Delta^{\alpha}(u) = \ln \bbT_{1, 1}( u + i/h ) - \ln \bbT_{1,1}(u_* + i/h ) =  \ln \bfT_{1, 1}( u + i/h ) - \ln \bbT_{1,1}(u_* + i/h ) -\mF(u + i/h )
$$
we find
\bea\label{eq:magic1}
\left[   
\ln{ \mathbb{ T }_{1, 1}( u_* \pm i/h ) }   \right]_{ 2 N } = 0, \hspace{1cm} N = 1, 2, \dots
\eea
This condition tells us that, when evaluated on a Riemann section defined with only ``short'' cuts of the form $(-2, +2) + in/h$, 
the function $\mathbb{ T }_{1, 1}$ has only two cuts with $\IIm(u) = \pm 1/h$. Because of this surprising property, this  Riemann section was called the ``magic sheet'' in \cite{Gromov:Solving}. Borrowing a further notation, we will denote with a hat $\hat{\text{T}}$ the analytic continuation of the 
T functions on a sheet with only short cuts, starting from their real values. Notice that, as discussed in Section \ref{sec2}, the Y-system and 
the T-system are naturally defined on a Riemann sheet with ``long'' branch cuts of the form $(-\infty, -2 ) \cup (+2, +\infty) + in/h$, a 
convention that is precisely the opposite of the ``magic sheet'' prescription.\\

The $\bbT$ gauge enjoys precisely the same analytic properties as the gauge denoted with the same font in \cite{Gromov:Solving}, describing the right band of the $\Bd$ diagram. This is not surprising, since the TBA equations relevant to describe the $(1, n)$ nodes are the same in $\Ad$ and $\Bd$. In particular, it turns out that, when evaluated on the magic sheet, the $ \hat{\mathbb{T}}_{a, s}$ functions with $s \geq a$ have at most two branch cuts each:
\begin{description}
\item[(a)]
$\hat{\mathbb{T}}_{1,n}$ has only two branch cuts on the magic sheet: $(-2,2)\pm in/h$ for $n \geq 1$
\item[(b)]
$\hat{\mathbb{T}}_{2,m}$ has only four branch cuts on the magic sheet: $(-2,2)\pm i(m-1)/h$, $(-2,2)\pm i(m+1)/h$ for $m \geq 2$
\item[(c)]
$\hat{\mathbb{T}}_{0,n} = 1$ for $n \geq -1$.
\end{description}
Moreover, the $\bbhat$ functions possess a special discrete symmetry, which in \cite{Gromov:Solving} was identified with a quantum version of the $\mathbb{Z}_4$ symmetry of the $\Bd$ sigma model. To discover this special symmetry one has to consider an extended domain given by the infinite horizontal band:
\bea
\label{eq:discrete}
\mathfrak{B} = \left\{ (a,s)\in \mathbb{Z}^2 \hspace{0.1cm} | \hspace{0.1cm} 0 \leq a \leq 2\right\}.
\eea
The solution on $\mathfrak{B}$ is constructed by assigning the values of $\bbhat$ to the nodes with $s \geq a$ and using the T-system on the magic sheet to populate the rest of the domain. Notice that, for $s > a$, the analyticity strips of the $\bbT$ functions are wide enough that the 
T-system equations hold even if we change $\bbT \rightarrow \bbhat$. However, this is no longer true at the nodes with $s = a$ and we expect that, 
for $s < a$, the T functions on $\mathfrak{B}$ bear no resemblance with the T functions on the corresponding nodes of the original $\Ad$ diagram.
 Therefore, for the sake of clarity we use the font $\barBB$ for the nodes with $s < a$ in $\mathfrak{B}$\footnote{ A good example is provided by 
the node $(1, 0)$. In $\mathfrak{B}$ we find $\barBB_{1, 0} = 0$, while in the original diagram there are \emph{two} functions $\bbhat_{1, 0}^{I}$,
 $\bbhat_{1, 0}^{II}$, and they are both different from zero according to (\ref{eq:FFFid}). }. With these notations, the special discrete symmetry 
can be written as follows:
\bea
\label{eq:Z4}
\barBB_{a, -s} = (-1)^a \hat{\mathbb{T}}_{a, s} , \hspace{1cm} s \geq a ,\hspace{2cm} \barBB_{a, -s} = (-1)^a \barBB_{a, s} ,\hspace{1cm} s < a . 
\eea
Although the proof of these properties can already be found in \cite{Gromov:Solving}, we find it useful to provide a partially alternative proof. We start by showing that
\bea\label{eq:asymptCond}
\bbhat_{2,2}(u) = \bbhat_{1,1}( u + 2i/h ) \bbhat_{1,1}( u -2i/h ).
\eea
One possible way to establish this result is notice that, as shown in the following subsection \ref{subapp1}, the ratio $G(u) = (  \bbhat_{1,1}( u + 2i/h ) \bbhat_{1,1}( u -2i/h ) )/ \bbhat_{2,2}(u) $ can be rewritten as the following combination of Y functions:
\bea\label{eq:GGG}
G(u) = \frac{( 1 + 1/Y_{2,2}(u + i/h ) )( ( 1 + 1/Y_{2,2}(u - i/h ) )  ( 1 + 1/Y_{2, 1}(u) )}{ Y_{1,0}^{I}(u_{\circl}) Y_{1,0}^{II}(u_{\circr})}.
\eea
Here, $u_{\circl}$ denotes the image of the point $u$ reached by analytic continuation through the branch cut with $\IIm(u) = +1/h$ and $u_{\circr}$, conversely, is the image of $u$ reached after following a path that encircles one of the branch points with $\IIm(u) = -1/h$. As we show in \ref{subapp2} below, from the TBA it is possible to prove that this quantity is precisely one, and this establishes (\ref{eq:asymptCond}).\\
Notice that, using the T-system on the nodes $(2, s)$ together with condition (\ref{eq:asymptCond}), this result can be generalised to $\bbhat_{2,s}(u) = \bbhat_{1,1}( u + si/h ) \bbhat_{1,1}( u -si/h )$ for $s \geq 1$. Since $\bbhat_{1,1}$ has only one pair of branch cuts, this proves property ${ \bf (b) }$.\\
To establish the remaining properties of the $\bbhat$ functions, let us derive some preliminary useful relations. Notice that (\ref{reflection}) implies $\bbT_{3, 2} =  \bbT_{2, 3} \mathcal{F}$, so that in the $\bbT$ gauge the identity $Y_{1,1} Y_{2,2} = ( T_{1,0}^I T_{1,0}^{II} T_{2,3})/(T_{0,1} T_{3,2})$ takes the form $ Y_{1,1} Y_{2,2} = ( \bbT_{1,0}^I \bbT_{1,0}^{II} )/(\mathcal{F})$. Comparing this result with (\ref{eq:discoF}), we find the important expression
\bea\label{eq:FFFid}
 \bbT_{1,0}^I(u) \bbT_{1,0}^{II}(u) = \mathcal{F}(u_* + i/h ).
\eea 
Finally, let us rewrite the discontinuity relations (\ref{eq:wDis}-\ref{eq:vDis}) in the $\bbT$ gauge. From the properties of the $\bbT$ functions listed above and using (\ref{eq:FFFid}), it is possible to show that (\ref{eq:vDis}) is equivalent to
\bea\label{eq:wDis2}
 \hat{\mathbb{T}}_{2, 2}( u - i/h )  \hat{\mathbb{T}}_{2, 2}( u + i/h )  =  \hat{\mathbb{T}}_{1, 1}( u - i/h )  \hat{\mathbb{T}}_{1, 1}( u + i/h ) \hat{\mathbb{T}}_{2, 3}(u).
\eea
The constraint (\ref{eq:wDis}) leads to the same equation but multiplied by a factor $\bbT_{1,3}( u + i/h )/(\bbT_{1,3}( u_* + i/h ))$ that we must set to one, therefore this provides another confirmation that $\bbT_{1,3} \in \mathcal{A}_3$.\\

Condition (\ref{eq:wDis2}) is very important to prove the symmetry property (\ref{eq:Z4}). In fact, it contains precisely the information needed to extend the solution from the right band into the left part of $\mathfrak{B}$.
Consider the T-system equation in $\mathfrak{B}$:
$\bbhat_{2,2}( u + i/h ) \bbhat_{2,2}( u - i/h ) = \bbhat_{2,3}(u) \barBB_{2,1}(u)$, where $\barBB_{2,1}(u)$ is so far unknown and defined by the previous relation. Comparing this equation with (\ref{eq:wDis2}) we find 
\bea
\label{eq:T21}
\barBB_{2, 1}(u) = \hat{\mathbb{T}}_{1, 1}( u - i/h )  \hat{\mathbb{T}}_{1, 1}( u + i/h ),
\eea
and matching (\ref{eq:T21}) with
$\hat{\mathbb{T}}_{1, 1}( u - i/h )  \hat{\mathbb{T}}_{1, 1}( u + i/h ) = \barBB_{1, 0}( u ) \bbhat_{1,2}(u) + 
\barBB_{2,1}(u) \bbhat_{0,1}(u)$ we get \bea \barBB_{1, 0}( u ) = 0 .\eea
Moreover, from (\ref{eq:T21}) we can compute 
$
\barBB_{2,1}( u + i/h ) \barBB_{2,1}( u - i/h ) = \bbhat_{1,1}^2(u) \bbhat_{1,1}(u + 2i/h)  \bbhat_{1,1}(u - 2i/h)
$. Using condition (\ref{eq:asymptCond}), we find that in order to match the T-system equation 
$\barBB_{2,1}( u + i/h ) \barBB_{2,1}( u - i/h ) = \barBB_{2,0}(u) \bbhat_{2, 2}(u)$ we have to take
\bea\label{eq:T20} 
\barBB_{2,0} = \bbhat_{1,1}^2 .
\eea 
Using (\ref{eq:T20}) and the T-system at the nodes $(1, 0)$ and $(2, 0)$ it is now easy to check that a solution constructed using the symmetry (\ref{eq:Z4}) satisfies the T-system at all nodes of $\mathfrak{B}$.\\
Finally, we refer the reader to a complex proof contained in \cite{Gromov:Solving}, Section ${\bf 4.2 }$. The authors show that a solution of 
the T-system on $\mathfrak{B}$ with the above mentioned properties including the discrete symmetry also has to satisfy
\bea
\left[ \ln\hat{\mathbb{T}}_{1,n} \right]_{ \pm( n + 2 m ) } = 0,\hspace{1cm} n , m = 1,2,\dots
\eea
Therefore, all $\bbhat_{1, s}$ have only two branch cuts. \\

This concludes the proof of the properties of the $\bbT$ functions. Including the discrete symmetry (\ref{eq:Z4}) and together with the properties (\ref{eq:bbTbasic}-\ref{eq:IequalsII}) of the $\bfT$ gauge, they are completely equivalent to the discontinuity relations. \\

Similarly to the $\Bd$ case, a discrete symmetry can also be derived for the $\bfhat$ functions. This symmetry emerges when considering the $\bfT$ functions on the magic sheet and extending them from the upper band to the following vertical domain:
\bea\label{eq:frakC}
\mathfrak{C} = \left\{ (n, s, \alpha) | \hspace{0.1cm} n \in \mathbb{Z}^2, s \in \left\{-1, 0\right\}, \hspace{0.1cm} \alpha \in
\left\{ I, II \right\} \right\}\hspace{0.2cm}\cup \hspace{0.2cm}\left\{ (n, s) | \hspace{0.1cm} n \in \mathbb{Z}, s\in\left\{ 1,2 \right\} \right\}.
\eea 
The original values $\bfhat_{n, s}$ are assigned to the nodes with $ n \geq s $ and the remaining T functions are computed by enforcing the validity of the T-system in the magic sheet kinematics in $\mathfrak{C}$. By very similar calculations as the ones reported above, one can construct a solution with the following symmetry:
\bea\label{eq:Z4double}
{ \hat{\bf T} }_{n, -1}^{\gamma} = - { \hat{\bf T} }_{-n, -1}^{\gamma} = 1, \hspace{0.3cm}
{ \hat{\bf T} }_{n, 0}^{\alpha} = { \hat{\bf T} }_{-n, 0}^{\beta}, \hspace{0.3cm}
{ \hat{\bf T} }_{n, 1}=-{ \hat{\bf T} }_{-n, 1}, \hspace{0.3cm}
{ \hat{\bf T} }_{n, 2} = { \hat{\bf T} }_{ -n, 2},
\eea
with $n \in \mathbb{Z}$, $\alpha$, $\beta$, $\gamma \in \left\{I, II\right\}$, $\alpha \neq \beta$. 
For simplicity of notation, we have used the same font $\bfhat_{n , s}$ for all the T functions in (\ref{eq:Z4double}). The reader should be aware that they differ from the T functions on the original diagram when $ n < s $.\\
Notice that there is a discontinuity in the first relation of (\ref{eq:Z4double}), 
and that the value of ${\hat{ \bf T} }_{0, -1}^{\gamma} $ appears 
double valued. In fact, the attentive reader will notice that T-system equations hold 
everywhere in $\mathfrak{C}$ apart from the nodes $(0 , -1, \gamma)$, $\gamma  \in \left\{I, II\right\}$. We can give an interpretation of this fact by viewing the extension of the T functions from the upper band to the whole of $\mathfrak{C}$ as an analytic continuation in their discrete indices.
In this case the analytic continuation introduces a branch point on each of the wings at the index value $(0, 0, \gamma)$, $\gamma = I, II$, and we trace the 
branch cuts to the left of these points, so that they cross the nodes $(0 , -1, \gamma)$.\\

As a last comment, let us compare the structure of these constraints with the ones found in \cite{Gromov:Solving} for $\Bd$. While the $\bbT$ gauge has precisely the same properties in the two systems, an important difference lies in the shape of the vertical domain on which the $\bfhat$ functions are endowed of their version of the discrete symmetry. In the $\Bd$ case, this was a strip $\left\{ -2, -1, 0, 1, 2\right\} 
\times \mathbb{Z}$, while in the present case it is given by $\mathfrak{C}$ defined in (\ref{eq:frakC}). It should be possible to derive 
FiNLIEs for the present case by adapting straightforwardly the methods of \cite{Gromov:Solving}. However one ingredient is still missing, 
namely finding a parametrization of the T-system on $\mathfrak{C}$ in terms of a finite number of Q functions.\\
Finally, while $\mathcal{F}$ plays in many respects the same r$\hat{\text{o}}$le here as in the $\Bd$ case, there is an important difference: in $\Bd$, single zeroes of $\mathcal{F}$ are interpreted as Bethe roots of the $\mathfrak{sl}_2$ sector, while in the present case we expect the Bethe roots to correspond to double zeroes of $\mF$. In fact, one can derive the expression 
\bea
\prod_{\alpha = I, II}( 1 + Y_{1,0}^{\alpha}(u) ) = \frac{\prod_{\alpha = I , II}\bbT_{1,0}^{\alpha}( u+i/h) \bbT_{1,0}^{\alpha}( u - i/h)}{ \bbT_{2,0}^I(u) \bbT_{2,0}^{II}(u)} = 
\frac{ \mF( u_{\circl} )\mF( u_{\circr}) }{ \bbT_{2,0}^I(u) \bbT_{2,0}^{II}(u) }
\eea
and by the analytic continuation $u\rightarrow u_{\circl}$ we find
\bea\label{eq:lastlast}
\prod_{\alpha= I, II}( 1 + Y_{1,0}^{\alpha}(u_{\circl}) ) = 
\frac{ \mF( u )\mF( (u_{\circr})_{\circl} ) }{ \prod_{\beta = I, II}\bbT_{2,0}^{\beta}(u) }.
\eea
Excited state TBA equations for the $\mathfrak{sl}_2$ subsector have been conjectured in \cite{GKVI,GKVII} for $\Bd$ and \cite{Levkovich} for $\Ad$. In our notation, the Bethe roots $u_j$ are described by the condition $Y_{1|\alpha}( (u_j)_{\circl} ) = -1$, therefore they are zeroes of the lhs of (\ref{eq:lastlast}). Because of the symmetry $Y_{1 , 0}^I = Y_{1, 0}^{II}$ in this sector and since $\mF(u) \neq \mF( (u_{\circr})_{\circl} ) $, we expect that $\mF$ exhibits a double zero. In $\Bd$, we would have the same expression but without the products over the $\alpha$, $\beta$ indices, thus leading to a single zero.

\subsection{ Proof of equation (\ref{eq:GGG}) }\label{subapp1}
Consider the identity 
$$( 1 + 1/Y_{1,1}(u) )= \frac{ \bbT_{1,1}( u + i/h ) \bbT_{1,1}(u-i/h) }{
\bbT_{1,0}(u) \bbT_{1,2}(u) }.
$$
After the analytic continuation $u \rightarrow u_*$ we get ( using the fact that $\bbT_{1,2}(u)$ has no branch points on the real axis
 and identity (\ref{eq:FFFid}))
$$
( 1 + Y_{2,2}(u) )= \frac{ \bbT_{1,1}( u_* + i/h ) \bbT_{1,1}(u_* - i/h) }{
 \mathcal{F}(u + i/h) \bbT_{1,2}(u) }.
$$
Shifting the previous expression starting from real $u$ allows us to reconstruct the product of $\bbhat_{1,1}( u + 2i/h ) \bbhat_{1,1}( u -2i/h )$.  Using $\bbT_{1,2}(u+i/h) \bbT_{1,2}(u-i/h) = ( 1 + Y_{1, 2}(u) ) \bbT_{2,2}(u)$ we arrive at ( for real $u$ )
$$
( 1 + Y_{2,2}(u + i/h ) )( ( 1 + Y_{2,2}(u - i/h ) )  ( 1 + Y_{1, 2}(u) ) = G(u) \frac{ \bbT_{1,1}(u_{\circl})
 \bbT_{1,1}(u_{\circr}) }{ \mathcal{F}^2(u) },
$$ 
where $u_{\circl}$ ( or $u_{\circr}$, respectively ) is the image of the point $u$ reached by analytic continuation through the branch cut with 
$\IIm(u) = +1/h$ ( resp. $\IIm(u) = -1/h$ ). Moreover using $Y_{1, 0}^{\alpha} = \frac{ \bbT_{1,1}}{ \bbT_{2, 0}^{\alpha} }$ we can rewrite the above identity as
$$
( 1 + Y_{2,2}(u + i/h ) )( ( 1 + Y_{2,2}(u - i/h ) )  ( 1 + Y_{1, 2}(u) ) = G(u) \frac{  Y_{1,0}^{I}(u_{\circl})}{ Y_{1,0}^{I}(u)} 
\frac{Y_{1,0}^{II}(u_{\circr}) }{  Y_{1,0}^{II}(u) } \frac{ \bbT_{1,1}^2(u)}{ \mathcal{F}^2(u) }
$$

$$
\frac{\mF( u_{\circl} )\mF( u_{\circr})}{\mF^2(u)} = (Y_{1,1}Y_{2,2}(u+i/h)) ( Y_{1,1}Y_{2,2}(u-i/h)).   
$$
At the same time we have:
$$
\prod_{\alpha}( 1 + 1/Y_{1,0}^{\alpha}(u) ) = \frac{\prod_{\alpha}\bbT_{1,0}^{\alpha}( u+i/h) \bbT_{1,0}^{\alpha}( u - i/h)}{ \bbT_{1,1}^2(u) } = 
\frac{ \mF( u_{\circl} )\mF( u_{\circr}) }{ \bbT_{1,1}^2(u) }.
$$
Putting all together and using the Y-system relation (\ref{Yysystem}) we arrive at (\ref{eq:GGG}).

\subsection{ Proof that $G(u) = 1$ }\label{subapp2}

We prove this relation starting from the TBA. We start by noticing that the relevant TBA kernels and the driving term satisfy the following identities:
\bea
\tilde{E}_{1}( u_{\circr} ) + \tilde{E}_{1}( u_{\circl} ) &=& 0 , \nn\\
\phi_{(y| \pm ),1}(z, u_{\circr} ) + \phi_{(y| \pm ),1}(z, u_{\circl} ) &=& \phi_1( z - u ), \nn\\
\phi_{(v|M),1}( z - u_{\circr}) + \phi_{(v|M),1}(z - u_{\circl}) &=& \phi_{M , 1}(z - u ), \nn\\
\phi_{(\CQ'|\alpha),(1|\beta)}(z , u_{\circr}) + \phi_{(\CQ'|\alpha),(1|\beta)}(z , u_{\circl}) &=& \phi_{\CQ',(v|1)}(z , u ), \forall \alpha, \beta \in \left\{I, II\right\}.
\eea
When applying the above analytic continuation to the convolutions in the TBA equation describing $\vep_{1}(u)$, some residue terms need to be
 taken into account. To list the relevant properties, let us give the following definitions:
\bea
A(u) &=& \int_{-2}^{2} dz\left( a(z)\,\phi_{(y|-),1}(z,u) - a(z_*)\,\phi_{(y|+),\CQ} (z,u)\right),\nn\\
B_M(u) &=&  \int_{-\infty}^{\infty} dz \, b(z) \phi_{(v|M),\CQ}(u), \nn\\ 
C_{\CQ,\alpha,\beta}(u) &=& \int_{-\infty}^{\infty} dz \, c(z) \phi_{(\CQ|\alpha),(1|\beta)}(z, u)\nn.
\eea
where $a$ denotes a function with two square root branch points at $u = \pm 2 $ and $b$, $c$ are two functions regular on the real axis. 
Then a careful monitoring of the movement of singularities leads to the following properties
\bea
A( u_{\circr} )  + A( u_{\circl} ) &=& \int_{-2}^{2} dz \left( a(z) - a(z_*) \right)\,\phi_{1}(z - u)  - a_+( u + i/h ) - a_+( u - i/h ), \nn\\
B_M( u_{\circr} )  + B_M( u_{\circl} ) &=& b*\phi_{M , 1}(u) - \delta_{M, 1} b(u) ,\nn\\
C_{\CQ,\alpha,\beta}( u_{\circr} )  + C_{\CQ,\alpha,\beta}( u_{\circl})  &=& \int_{-\infty}^{\infty} dz \, c(z) \phi_{(\CQ|\alpha),(v|1)}(z, u) ,\nn
\eea
where $a_+(u) = a(u_*)$.\\
Using these relations, from the TBA equation describing $\vep_{1}(u)$ we obtain
\bea
\vep_1( u_{\circr} )  +  \vep_1( u_{\circl} ) = -\sum_{\beta}\sum_{\CQ'=1}^{\infty}L_{\CQ'|\beta}*\phi_{(\CQ|\beta),(v|1)}(u)
+\sum_{M=1}^{\infty}L_{v|M}*\phi_{M,1}(u) \nn \\ 
+\int_{-2}^{2} dz\left[L_{y|-}(z)
- L_{y|+}(z)\right]\phi_1(z-u)\nn\\
- L_{y|+}( u + i/h ) - L_{y|+}( u - i/h ) -L_{v|1}(u)
\eea
and by comparison with the TBA equation for $\vep_{v|1}(u)$ this can be rewritten as
\bea
\vep_1( u_{\circr} )  +  \vep_1( u_{\circl} ) = 
- L_{y|+}( u + i/h ) - L_{y|+}( u - i/h ) - \ln( 1 + Y_{v|1}(u) ).
\eea
This is precisely the statement that $G(u) = 1$.


\end{document}